\documentclass[conference]{IEEEtran}
\IEEEoverridecommandlockouts
\usepackage{amsmath,amsfonts}
\usepackage{algorithmic}
\usepackage{array}
\usepackage[caption=false,font=normalsize,labelfont=sf,textfont=sf]{subfig}
\usepackage{textcomp}
\usepackage{stfloats}
\usepackage{url}
\usepackage{verbatim}
\usepackage{graphicx}
\def\BibTeX{{\rm B\kern-.05em{\sc i\kern-.025em b}\kern-.08em
    T\kern-.1667em\lower.7ex\hbox{E}\kern-.125emX}}
\usepackage{balance}
\usepackage{nicematrix}
\usepackage{multirow}
\usepackage{enumerate}
\usepackage[utf8]{inputenc}
\usepackage{graphicx}
\usepackage{float}
\usepackage{cite}
\usepackage{hyperref}
\usepackage{adjustbox}
\usepackage[skip=2pt]{caption} 
\usepackage{enumitem}
\usepackage{booktabs}
\usepackage{caption}
\usepackage{fancyhdr} 
\usepackage{xcolor}
\usepackage{algorithm}

\fancypagestyle{myfooter}{
  \fancyfoot[C]{\href{链接地址}{超链接内容}} 
}
\begin{document}

\title{Intent-guided Heterogeneous Graph Contrastive Learning for Recommendation}

\author{
\IEEEauthorblockN{
Lei Sang\textsuperscript{1}, 
Yu Wang\textsuperscript{1}, 
Yi Zhang\textsuperscript{1}, 
Yiwen Zhang\textsuperscript{1*}, 
Xindong Wu\textsuperscript{2}}
\IEEEauthorblockA{\textsuperscript{1}\textit{Anhui University}, Hefei, China \\
sanglei@ahu.edu.cn, wangyuahu@stu.ahu.edu.cn, zhangyi@stu.ahu.edu.cn, zhangyiwen@ahu.edu.cn}
\IEEEauthorblockA{\textsuperscript{2}\textit{Hefei University of Technology},  Hefei, China \\
xwu@hfut.edu.cn}
\thanks{* Corresponding author is Yiwen Zhang.}
}
\maketitle

\begin{abstract}
Contrastive Learning (CL)-based recommender systems have gained prominence in the context of Heterogeneous Graph (HG) due to their capacity to enhance the consistency of representations across different views. However, existing frameworks often neglect the fact that user-item interactions within HG are governed by diverse latent intents (e.g., brand preferences or demographic characteristics of item audiences), which are pivotal in capturing fine-grained relations. The exploration of these underlying intents, particularly through the lens of meta-paths in HGs, presents us with two principal challenges: i) How to integrate CL with intents; ii) How to mitigate noise from meta-path-driven intents.

To address these challenges, we propose an innovative framework termed \textit{Intent-guided Heterogeneous Graph Contrastive Learning} (IHGCL), which designed to enhance CL-based recommendation by capturing the intents contained within meta-paths. Specifically, the IHGCL framework includes: i) a meta-path-based Dual Contrastive Learning (DCL) approach to effectively integrate intents into the recommendation, constructing intent-intent contrast and intent-interaction contrast; ii) a Bottlenecked AutoEncoder (BAE) that combines mask propagation with the information bottleneck principle to significantly reduce noise perturbations introduced by meta-paths. Empirical evaluations conducted across six distinct datasets demonstrate the superior performance of our IHGCL framework relative to conventional baseline methods. Our model implementation is available at https://github.com/wangyu0627/IHGCL.
\end{abstract}


\section{Introduction}
Recommender systems \cite{2010rs, 2023bsl} play an increasingly crucial role in daily life, including content delivery in short videos \cite{2020cdprec}, news \cite{2023news}, and shopping \cite{2020lightgcn}. These systems use users' implicit historical interactions to effectively assist them in discovering items or products that align with their preferences \cite{2012bpr}. Traditional recommendation methods \cite{2019ngcf, 2020lightgcn} often overlook the potential value of users' auxiliary attributes and items' label features. These methods rely solely on collaborative filtering \cite{2017ncf, 2019ngcf} through the user-item interaction graph \cite{2020lightgcn} to infer user preferences. However, in many real-world scenarios, the available interaction data are typically highly sparse, posing challenges for these traditional approaches.

\begin{figure}[t]
    \centering
    \includegraphics[width=\linewidth]{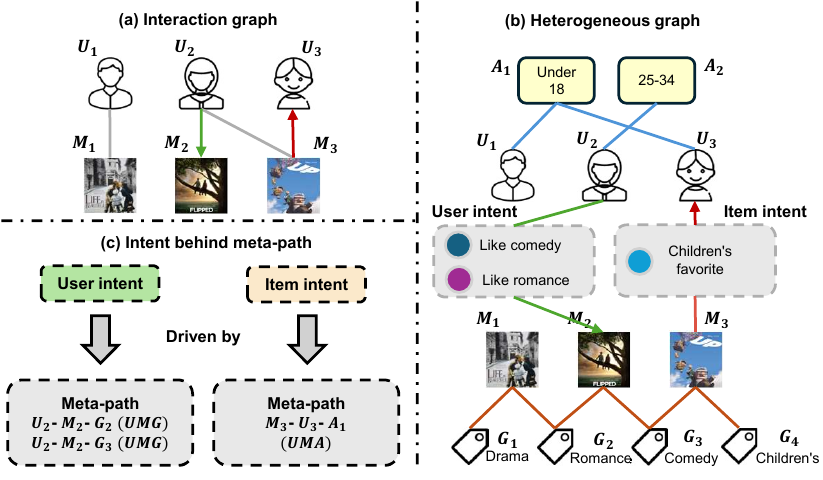}
    \caption{(a) Interaction graph in a movie scenario, where the green arrow indicates recommending movie $M_2$ to user $U_2$, and the red arrow indicates user $U_3$ to movie $M_3$; (b)  heterogeneous graph incorporating user and item intents, showing that interactions are guided by intents; (c) considering the intents driven by the meta-paths.}
    \label{fig1:CH1}
    \vspace{-0.8cm}
\end{figure}

Contrastive learning (CL)-based recommendation approaches have emerged as a novel perspective to address the limitations of traditional recommendation methods \cite{2020simclr, 2021sgl, 2020moco}. These approaches aim to maximize the consistency of representations across different views, in contrast to the conventional collaborative filtering techniques. CL-based recommendation models \cite{2022directau, 2022simgcl, 2021sgl, 2023dccf} employ the principles of alignment and uniformity \cite{2022directau} to spread node embeddings apart and maximize the information entropy in the embedding space. This ensures that similar nodes are positioned closely together, while dissimilar nodes are distant from each other. In essence, the embeddings of users and items should be distributed in the space in a tight and dispersed manner. For instance, SimGCL \cite{2022simgcl} and SGL \cite{2021sgl} utilize Gaussian noise and graph augmentation, respectively, to implement this concept. Building upon these advancements, CL-based approaches have been further adapted to the more complex heterogeneous graph (HG) recommendation scenarios \cite{2013hin, 2020cdprec, 2018herec}, where multiple types of nodes and relationships are modeled. For example, HGCL \cite{2023hgcl} constructs contrastive views by combining HG-based auxiliary information with user-item interactions. These CL-based recommendation models have demonstrated promising performance in various settings.

Despite their effectiveness and explainability, to the best of our knowledge, these studies largely overlook the underlying fine-grained intents of users and items. Taking Fig. \ref{fig1:CH1} as an example, the part (a) represents the bipartite graph recommendation paradigm, while the Fig. \ref{fig1:CH1} (b) illustrates the modeling user and item intents from existing information. Meta-paths \cite{2017metapath2vec, 2019han} are commonly used in heterogeneous graph as tools for capturing preferences between nodes. For example, the intent between user $U_3$ and movie $M_2$ is driven by the meta-paths `$U_2-M_2-G_2$' and `$U_2-M_2-G_3$'. We then consider recommending $M_1$ (belongs to $G_2$), or $M_3$ (belongs to $G_3$), to user $U_2$. Similarly, the intent between movie $M_3$ and user $U_3$ is the meta-path `$M_3-U_3-A_1$'. In this case, recommending movie watched by user $U_1$ (genre $A_1$) is more suitable. These observations underscore the profound influence of respective intents on both users and items. Heterogeneous Graph Neural Network (HGNN) \cite{2021heco, 2019han} can effectively capture richer and deeper behaviors and preferences of users and items from complex data structures by meta-path based aggregation. This motivates us to employ this approach to capture the intents behind meta-paths to enhance CL-based recommendation. However, we still face the following two challenges:

\textbf{C1: How to integrate CL with intents?}
Existing heterogeneous graph recommendation methods \cite{2022hgcl, 2023hgcl, 2021heco} have investigated the effectiveness of contrastive learning in transferring information. These methods design contrastive objectives between the auxiliary heterogeneous information and the main user-item view. However, they have not constructed  multiple contrastive views based on the intents behind meta-paths. Multi-view contrasts have been proven to be vital in improving recommendation performance \cite{2021sgl, 2023lightgcl, 2024graphaug}. The rich semantic information from meta-paths contains more fine-grained user preferences and item features, which are powerful tools for capturing intents. Overall, modeling users' and items' intents and integrating them into the main user-item view to construct contrastive objectives is a meaningful challenge.

\begin{figure}[t]
    \centering
    \begin{minipage}[b]{0.49\linewidth}
        \centering
        \subfloat[\small Douban Movie \cite{2018herec}]{\includegraphics[width=\linewidth]{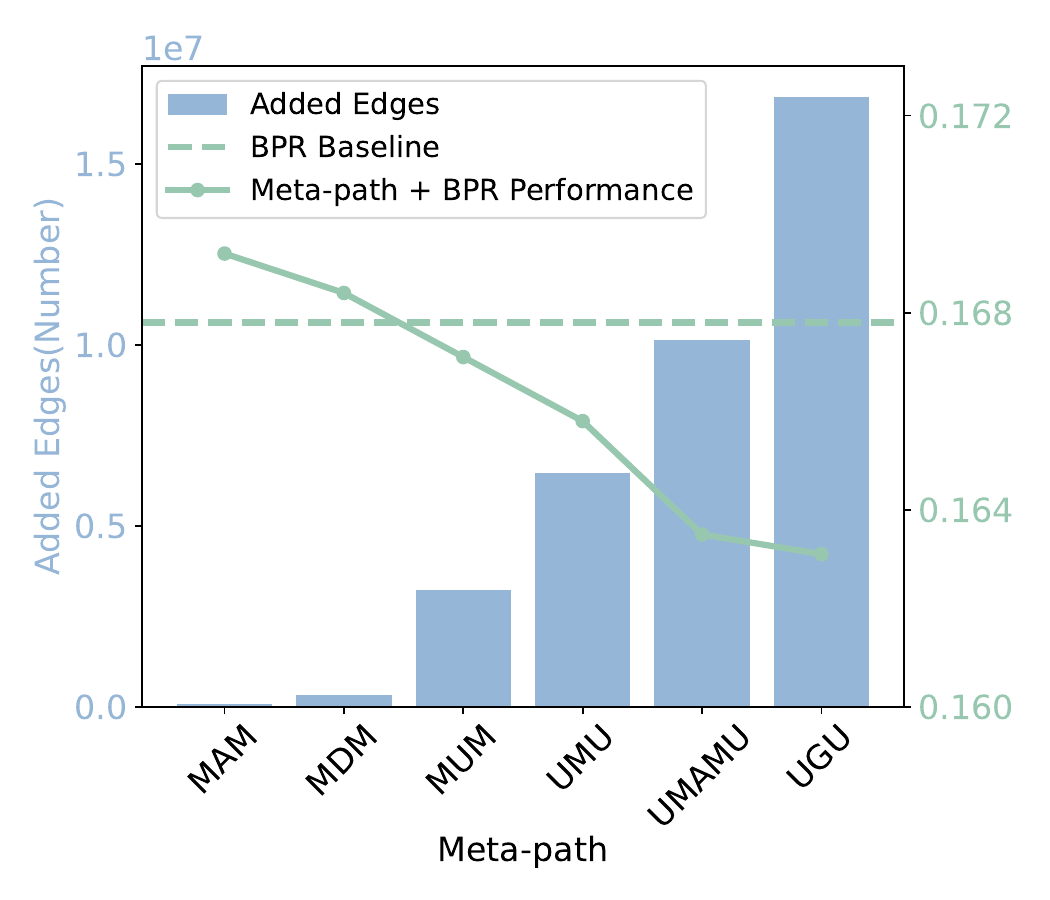}}
    \end{minipage}
    \begin{minipage}[b]{0.49\linewidth}
        \centering
        \subfloat[\small Yelp \cite{2023chest}]{\includegraphics[width=\linewidth]{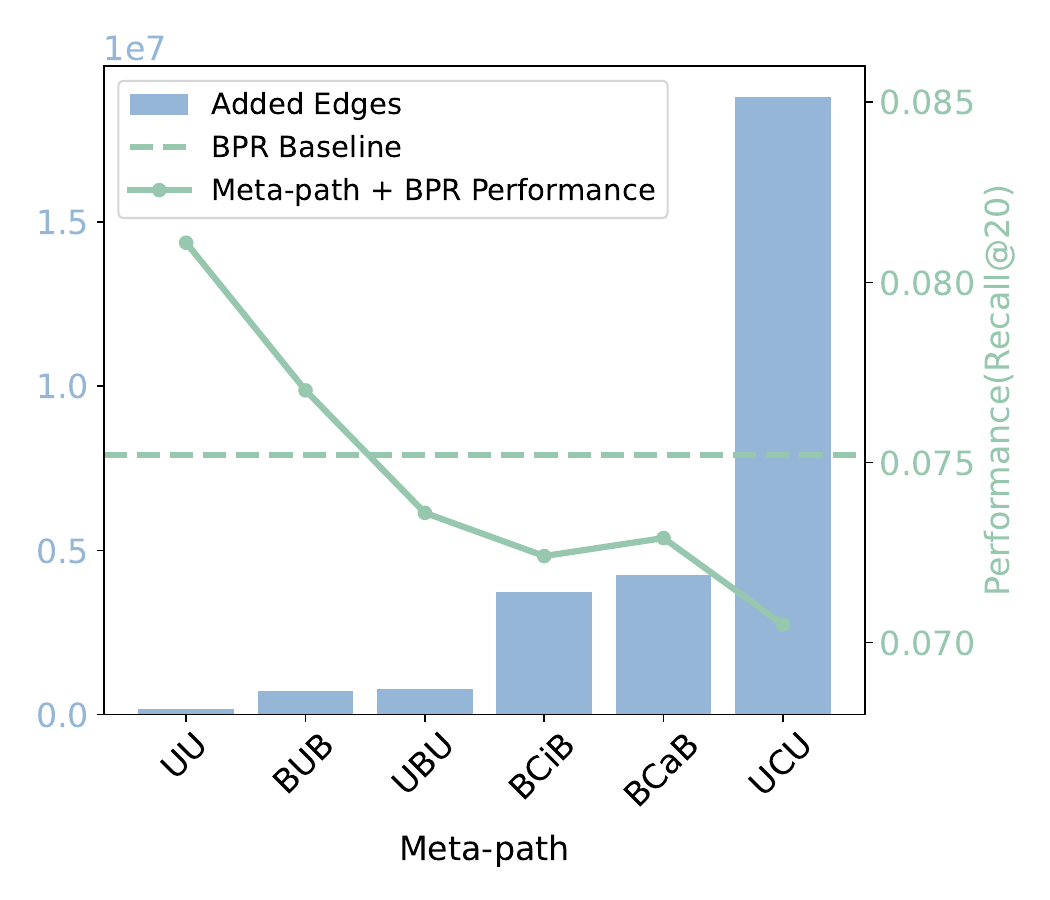}}
    \end{minipage}
    \caption{The impact w.r.t. different meta-paths. The blue bar is the number of added edges current meta-path, and the green line indicates the performance for current meta-path.}
    \label{fig2:CH2}
    \vspace{-0.8cm}
\end{figure}

\textbf{C2: How to mitigate noise from meta-path-driven intents?}
In heterogeneous graph, intent-based meta-paths are used to construct new connections between users or items, which help alleviate data sparsity and enhance recommendation performance. However, it is challenging to distinguish whether these connections are informative or merely noise. As illustrated in Fig. \ref{fig2:CH2}, we construct user-user or item-item interaction graphs using a single meta-path at each time to improve recommendation under the BPR \cite{2012bpr} loss. Experimental results demonstrate that when a specific type of meta-path introduces more connections, it may also lead to a decline in recommendation performance. Therefore, we hypothesize that although meta-paths bring rich semantic information, they also contain indescribable noise \cite{2023post_train, 2021MvDGAE}. This noise propagated through the layer of HGNN to farther nodes diminishes the recommendation performance.

To tackle the challenges mentioned above, we propose an Intent-guided Heterogeneous Graph Contrastive Learning (\textbf{IHGCL}) framework for recommendation. 
First, for \textbf{C1}, A meta-path enhanced \textbf{Dual Contrastive Learning} (DCL) is proposed, which aims to align the different intent-based embeddings learned by users and items. Intent-intent contrast unifies user and item preferences by maximizing the consistency of intent embeddings. Intent-interaction contrast involves adding intent embeddings to the representation of real interactions to perform representation-level data augmentation.
For \textbf{C2}, we design an \textbf{Bottlenecked Autoencoder} (BAE) to mitigate the noise issues induced by meta-paths. BAE reconstructs robust and denoised node representations via a dual-masked autoencoder, with the masking operation applied to the node embeddings. It also introduces an information bottleneck loss to constrain the amount of information between the autoencoder and the graph structure, aiming to find the \textbf{minimum sufficient} representation of a dataset. 
Overall, IHGCL is an effective framework for HG-based recommendation that considers the potential of intents behind meta-paths for constructing contrastive views and effectively mitigates the noise issues introduced by intents. The contributions are summarized as follows:

\begin{itemize}[leftmargin=*]
\item We explore the intents behind meta-paths and integrate them into contrastive learning to construct views, and propose the recommendation framework IHGCL, which models the intents of users and items via a Dual Contrastive Learning (DCL) module. 
\item We further propose an Bottlenecked Autoencoder (BAE), which effectively mitigates the noise issues introduced by the meta-paths, while adaptively preserving the topological structure through information bottleneck techniques.
\item We conduct extensive experiments on six public datasets, and the results show that IHGCL not only outperforms existing baselines but also demonstrate effectiveness from various aspects.
\end{itemize}

\section{Preliminaries} \label{sec2.1}
\subsection{Definitions}
In this section, we formally define some significant concepts related to heterogeneous graph as follows:

\textbf{Heterogeneous Graph (HG \cite{2019han}):} A HG is characterized by a graph structure $\mathcal{G}=(\mathcal{V}, \mathcal{E}, \mathcal{A}, \mathcal{R}, \phi, \varphi)$, where the collections of nodes and edges are symbolized by $\mathcal{V}$ and $\mathcal{E}$, respectively. For each node $v$ and edge $e$, associated type mapping functions exist: $\phi: \mathcal{V} \rightarrow \mathcal{A}$ for nodes and $\varphi: \mathcal{E} \rightarrow \mathcal{R}$ for edges. Here, $\mathcal{A}$ represents the node types and $\mathcal{R}$ denotes the edge types, with the total number of types exceeding two, i.e., $|\mathcal{A}|+|\mathcal{R}|>2$.

\begin{figure}[t]
    \centering
    \includegraphics[width=0.98\linewidth]{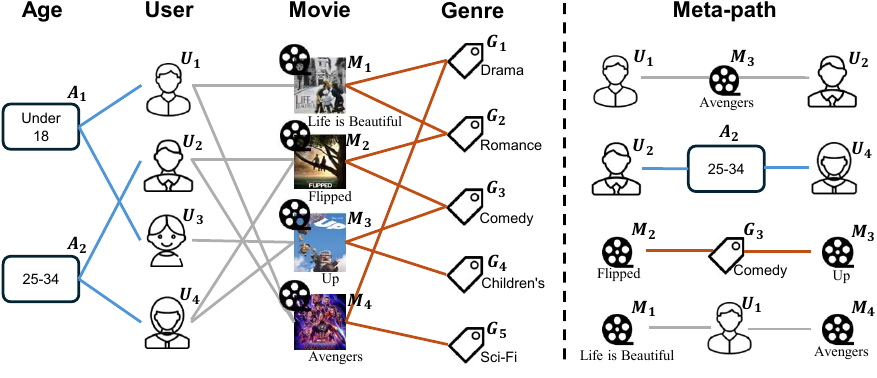}
    \caption{A toy example of heterogeneous graph on Movielens \cite{zhang2023revisiting} dataset for recommendation.}
    \label{fig3:preliminaries}
    \vspace{-0.5cm}
\end{figure}

\textbf{Meta-path \cite{2017metapath2vec}:} In a $\mathcal{G}$, a meta-path $\rho$ is represented as $\mathcal{A}_{1} \stackrel{\mathcal{R}_{1}}{\longrightarrow} \mathcal{A}_{2} \stackrel{\mathcal{R}_{2}}{\longrightarrow} \ldots \stackrel{\mathcal{R}_{l}}{\longrightarrow} \mathcal{A}_{l+1}$, describing a composite connection between $\mathcal{A}_{1}$ and $\mathcal{A}_{l+1}$. Figure \ref{fig3:preliminaries} shows that we establish connections between users (or items) from the heterogeneous graph on the left according to the given meta-paths. For instance, users $U_{1}$ and $U_{2}$ can be connected by the path ``$U_{1}-M_{3}-U_{2}$'' to enrich the data. Typically, different meta-paths reveal varied dependency information between two nodes. The path ``$UAU$'' indicates that two users belong to the same age group, while ``$MGM$'' suggests that two movies belong to the same genre.

\textbf{Meta-path-based Subgraph:} We define a specific meta-path $\rho \in \mathcal{R}$ and the corresponding node type $V \in \mathcal{A}$. For a node $v \in V$, the set of edges formed by connecting $v^\rho$ to new nodes through the meta-path $\rho$ are denoted as $e$. By traversing other nodes in $V$, we obtain a set of edges $E$. Consequently, $E$ and the set of all nodes of this specific type (denoted by $V$) constitute a subgraph $G_V^\rho$. Taking Figure \ref{fig3:preliminaries} as an instance, we define $\rho$ as ``$MGM$'' and the node type $V$ as ``movie''. All movie-movie connections formed through this meta-path constitute the meta-path-based subgraph $G_M^{MGM}$.

\subsection{Problem Formulation} \label{sec2.2}
Typical recommendation scenarios include a set of $M$ users $\mathcal{U}=\left\{u_{1}, u_{2}, \ldots, u_{M}\right\}$ and a set of $N$ items $\mathcal{I}=\left\{i_{1}, i_{2}, \ldots, i_{N}\right\}$. Furthermore, historical user-item interaction records are stored in a matrix $\mathbf{R}^{M \times N}$. Some research \cite{2019ngcf, 2020lightgcn} define a bipartite graph structure:
\begin{equation}
    \mathbf{A}=\left(\begin{array}{cc}
    \mathbf{0} & \mathbf{R} \\
    \mathbf{R}^{\top} & \mathbf{0}
\end{array}\right),
\label{eq1}
\end{equation}

Let the initialized embedding be $\mathbf{E}^{(0)} \in \mathbb{R}^{(M+N) \times d}$, where $d$ is the embedding dimension. $\mathbf{E}^{(0)}$ includes $\mathbf{E}_u^{(0)}$ and $\mathbf{E}_i^{(0)}$, and through normalization propagation \cite{2020lightgcn}, we obtain the $l$-th layer embeddings: 
\begin{equation}
    \mathbf{E}^{(l+1)}=\left(\mathbf{D}^{-0.5} \mathbf{A} \mathbf{D}^{-0.5}\right) \mathbf{E}^{(l)},
\label{eq2}
\end{equation}
where $\mathbf{D}$ is a degree matrix used to measure the number of non-zero entries in each row of $\mathbf{A}$. The existing methods for contrastive learning can generally be categorized into data-based (DA), feature-based (FA), and model-based (MA) augmentations \cite{2024sslrec}:
\begin{equation}
\left\{
\begin{aligned}
    \text{DA:} &\quad \mathbf{A}^{'} = \mathcal{T}^{'}(\mathbf{A}), \quad \mathbf{A}^{''} = \mathcal{T}^{''}(\mathbf{A}) \\
    \text{FA:} &\quad \mathbf{E}^{'} = \mathbf{E} + \Delta^{'}, \quad \mathbf{E}^{''} = \mathbf{E} + \Delta^{''} \\
    \text{MA:} &\quad \mathbf{Z}^{'} = \mathcal{F}^{'}(\mathbf{A}, \mathbf{E}), \quad \mathbf{Z}^{''} = \mathcal{F}^{''}(\mathbf{A}, \mathbf{E})
\end{aligned}
\right.
\end{equation}
where $\mathcal{T}$ is the graph structure transformer, such as random edge dropout and structural learning. $\Delta$ typically represents noise with a specified distribution. $\mathcal{F}$ is defined as a view generator containing learnable parameters. For example, an intent disentanglement module used to capture user intent. The common practice in CL-based recommendation \cite{2021sgl, 2022simgcl, 2023lightgcl} involves generating two augmented views combined with the main view for contrast, which goes against the existing paradigm \cite{2022hgcl, 2023hgcl} of heterogeneous contrastive recommendation (aligning auxiliary information with the main task). We delve into this paradigm in our motivation and propose a
solution.

\begin{figure*}[t]
    \centering
    \includegraphics[width=\linewidth]{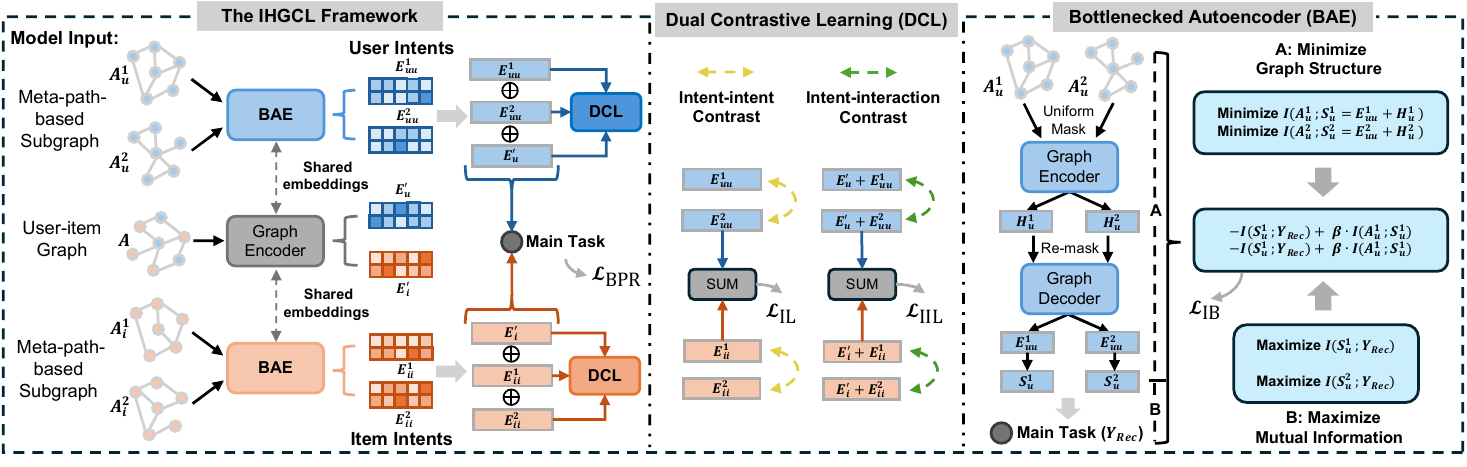}
    \caption{The complete framework of the proposed IHGCL, which consists of a Dual Contrastive Learning (DCL) module and a Bottlenecked Autoencoder (BAE). The DCL module generates two types of contrasts: contrasts between meta-paths and contrasts between meta-path-enhanced views, providing self-supervised signals. The BAE module employs a dual-masked autoencoder combined with an adaptive information bottleneck technique to mitigate the noise issues, which can capture the \textbf{minimum sufficient information} from the data features.}
    \label{fig4:model}
    \vspace{-0.5cm}
\end{figure*}

\section{METHODOLOGY}
The overview of the proposed IHGCL is shown in Figure \ref{fig4:model}. We propose a model-based augmentation for contrastive learning, which leverages the intents in heterogeneous information to model the user and item intents. The model comprises into four modules: model input, an bottlenecked autoencoder, dual contrastive learning, and model optimization.

\subsection{\textbf{Model Input}}
The input to the model is divided into the main user-item view and the auxiliary heterogeneous information from the user view and item view.
\begin{itemize}[leftmargin=*]
\item \textbf{User-item view.} We follow the classic CL-based recommendation task \cite{2021sgl, 2022simgcl}, using $\mathbf{A}$ mentioned in Section \ref{sec2.2} and embeddings $\mathbf{E}_u$ and $\mathbf{E}_i$ for users and items as the main view's input. $\mathbf{E}_u$ and $\mathbf{E}_i$ serve as shared parameters in the parallel training process.
\item \textbf{Heterogeneous view.} To construct multi-view augmentations for contrastive learning, we model the preferences of user $u$ and attributes of item $i$, which can be achieved through meta-paths \cite{2018herec, 2023hgcl}. A specific meta-path that connects two nodes embodies a similar semantic relation, that is, the user's group and the item's category. Based on the Section \ref{sec2.1}, we respectively select two meta-path-based subgraphs of users and items for the model: 
\begin{equation}
    \mathbf{User}: \boldsymbol{G}_U^{\rho^u_1}, \boldsymbol{G}_U^{\rho^u_2}; \quad \mathbf{item}: \boldsymbol{G}_I^{\rho^i_1}, \boldsymbol{G}_I^{\rho^i_2},
    \label{eq4}
\end{equation}
where $\rho^u_k$ denotes the $k$-th meta-path based on users, and the items are the same definition.
\end{itemize}
The five graphs on the left in Figure \ref{fig4:model}, along with the embeddings of users and items, constitute the input of the model. Such an operation often brings a substantial amount of interaction data to alleviate sparsity in recommendation. However, we have demonstrated in Figure \ref{fig2:CH2} that such subgraphs usually contain noise \cite{2023post_train, 2021MvDGAE}.

\subsection{\textbf{Bottlenecked Autoencoder (BAE)}} \label{2.3.2}
In this section, we introduce the BAE to model rich intents and mitigate the noise in heterogeneous information. The masked autoencoder can better reconstruct noise-polluted data by learning the global structure of the data. The information bottleneck method improves reconstruction accuracy by retaining \textbf{minimum sufficient} information to the recommendation.

Given the rich intents from meta-paths, we need to preserve the user preferences and item attributes to the greatest extent without randomly disrupting the data structure. Many studies \cite{2023hgcml, 2021heco} have demonstrated that the similar intents of each meta-path can be treated as a distinct view, and for each view, we employ the same BAE. First, the model defines meta-path-based subgraphs of users and items as: 
\begin{equation}
    \boldsymbol{G}_U^j = (\boldsymbol{V}_u, \mathbf{A}_u^j, \mathbf{E}_u), \quad
    \boldsymbol{G}_I^j = (\boldsymbol{V}_i, \mathbf{A}_i^j, \mathbf{E}_i) \quad \text{for} \; j = 1, 2
\label{eq5}
\end{equation}
where $\mathbf{A}_u^j$ and $\mathbf{A}_i^j$ are the matrices corresponding to Eq. \ref{eq4}, and $j$ represents the $j$-th meta-path of user or item. $|\boldsymbol{V}_u|=M$ and $|\boldsymbol{V}_i|=N$ are the number of users and items. Next, we employ a sampling strategy without replacement to obtain the node set $\boldsymbol{V}^{[Mask]}$ with masks before entering the encoder:
\begin{equation}
\begin{aligned}
    \boldsymbol{V}^{[Mask]}_u &= \{ \boldsymbol{V}_{s} \in \boldsymbol{V}_{u} \mid r_s \leq p \}, & r_s &\sim \text{Uniform}(0, 1), \\
    \boldsymbol{V}^{[Mask]}_i &= \{ \boldsymbol{V}_{s} \in \boldsymbol{V}_{i} \mid r_s \leq p \}, & r_s &\sim \text{Uniform}(0, 1),
\end{aligned}
\end{equation}
where $p$ is the mask ratio, and $\boldsymbol{V}_{s}$ is the node set based on $r_s$. To mitigate the risk of sampling biases, where a node's entire neighbor is either fully masked or fully visible, uniform random sampling is employed. This technique enhances the encoder's generalization ability by avoiding localized bias centers. We mask the entire node embeddings in the set $\boldsymbol{V}^{[Mask]}$ to obtain $\widetilde{\mathbf{E}}_{u} = \{ \mathbf{e}_u^1, \mathbf{e}_u^2, \dots, \mathbf{e}_u^{M} \}$ and $\widetilde{\mathbf{E}}_{i} = \{ \mathbf{e}_i^1, \mathbf{e}_i^2, \dots, \mathbf{e}_i^{N} \}$. Next, we empirically employ LightGCN \cite{2020lightgcn} as the encoder to perform convolution on the masked embeddings:
\begin{equation}
\begin{aligned}
    \mathbf{h}_u^{(L_E)} &= \sum_{v \in \mathcal{N}u} \frac{1}{\sqrt{|\mathcal{N}_u||\mathcal{N}_v|}} \mathbf{h}_v^{(L_E-1)}, \\ 
    \mathbf{h}_i^{(L_E)} &= \sum_{j \in \mathcal{N}_i} \frac{1}{\sqrt{|\mathcal{N}_i||\mathcal{N}_j|}} \mathbf{h}_j^{(L_E-1)},
\end{aligned}   
\end{equation}
where $L_E$ is the encoder layer, and $\mathbf{h}_v^{(0)}$ and $\mathbf{h}_j^{(0)}$ denote one of the nodes $\widetilde{\mathbf{E}}_{u}$ and $\widetilde{\mathbf{E}}_{i}$. $\mathcal{N}_u$ and $\mathcal{N}_i$ represent the first-order receptive fields of users and items, respectively. The first reconstructed embeddings are $\mathbf{H}_u = \{ \mathbf{h}_u^1, \mathbf{h}_u^2, \dots, \mathbf{h}_u^{M} \}$ and $\mathbf{H}_i = \{ \mathbf{h}_i^1, \mathbf{h}_i^2, \dots, \mathbf{h}_i^{N} \}$. From a macro perspective, this is a matrix multiplication operation: $\left(\mathbf{D}_{u}^{-0.5} \mathbf{A}_{u} \mathbf{D}_{u}^{-0.5}\right)^{L_E} \widetilde{\mathbf{E}}_{u}$ and $\left(\mathbf{D}_{i}^{-0.5} \mathbf{A}_{i} \mathbf{D}_{i}^{-0.5}\right)^{L_E} \widetilde{\mathbf{E}}_{i}$, where $\mathbf{D}_{u}$ and $\mathbf{D}_{i}$ are the diagonal degree matrices of $\mathbf{A}_{u}$ and $\mathbf{A}_{i}$ ($\mathbf{A}_u \in \mathbb{R}^{M \times M}$ and $\mathbf{A}_i \in \mathbb{R}^{N \times N}$). 

Considering that nodes aggregate information from a subset of their neighbors, these operations reduce dependency on specific nodes. However, when the masking rate is low, the node embeddings may still contain direct information from the original input features (i.e., noise), leading to the failure of model-based augmentations. Therefore, we adopt the re-masking and decoder to address this issue. BAE employs the same sampling strategy to obtain a new set $\boldsymbol{V}^{[Remask]}$ for masking the reconstructed embeddings to obtain $\widetilde{\mathbf{H}}_{u}$ and $\widetilde{\mathbf{H}}_{i}$. The construction of the encoder and decoder is identical, except that the input embeddings are re-masked: 
\begin{equation}
\begin{aligned}
    \mathbf{E}_{uu} &= \left(\mathbf{D}_{u}^{-0.5} \mathbf{A}_{u} \mathbf{D}_{u}^{-0.5}\right)^{L_D} \widetilde{\mathbf{H}}_{u}, \\
    \mathbf{E}_{ii} &= \left(\mathbf{D}_{i}^{-0.5} \mathbf{A}_{i} \mathbf{D}_{i}^{-0.5}\right)^{L_D} \widetilde{\mathbf{H}}_{i},
\end{aligned}
\label{eq8}
\end{equation}
where $L_D$ is the decoder layer. $\mathbf{E}_{uu}$ and $\mathbf{E}_{ii}$ are the final outputs of BAE. Expanding to multiple heterogeneous views and based on different inputs, we can obtain $\mathbf{E}_{uu}^1$, $\mathbf{E}_{uu}^2$, $\mathbf{E}_{ii}^1$ and $\mathbf{E}_{ii}^2$.

We use GNNs \cite{2019ngcf, 2022vibgsl} as encoders and decoders due to their strong capability to aggregate neighborhood information and iteratively propagate it. Moreover, LightGCN typically performs this task well, thereby enhancing the efficiency of recommendation. However, such autoencoders overly rely on the masked nodes, which leads to insufficient generalization capability. Thus, we apply the Information Bottleneck (IB) \cite{2016dvib, 2022cgi} to constrain the encoder's and decoder's embeddings, which limits the amount of information between the autoencoder and the graph structure, thereby forcing the model to focus on the essential features of the data. This method enhances the model's robustness to noise. First, we adopt the loss function $\mathcal{L}_{\text{IB}}$ between the autoencoder's embeddings and graph structure. Specifically, we define $\left(\mathbf{E}_{uu} + \mathbf{H}_u\right)$ and $\left(\mathbf{E}_{ii} + \mathbf{H}_i\right)$ as $\mathbf{S}_u$ and $\mathbf{S}_i$, respectively. The information bottleneck losses are as the follow:
\begin{equation}
\begin{aligned}
\mathcal{L}_{\text{UIB}}(\mathbf{S}_u^j, \mathbf{A}_u^j; \mathbf{Y}_{Rec}) &= - I(\mathbf{S}_{u}^j; \mathbf{Y}_{Rec}) + \beta \cdot I(\mathbf{A}_u^j; \mathbf{S}_{u}^j), \; j = 1, 2 \\
\mathcal{L}_{\text{IIB}}(\mathbf{S}_{i}^j, \mathbf{A}_i^j; \mathbf{Y}_{Rec}) &= - I(\mathbf{S}_{i}^j; \mathbf{Y}_{Rec}) + \beta \cdot I(\mathbf{A}_i^j; \mathbf{S}_{i}^j), \; j = 1, 2
\end{aligned}
\label{eq9}
\end{equation}
where $I(\cdot; \cdot)$ is the mutual information function, and $\beta$ is a non-negative Lagrangian multiplier \cite{2016dvib} employed to control the compression strength. $j$ denotes different meta-paths (Eq. \ref{eq5}). The $\mathbf{Y}_{Rec}$ is supervised signal \cite{2022cgi}, which in recommendation task corresponds to BPR \cite{2012bpr} interaction pairs. However, $\mathbf{A}_{u}$ or $\mathbf{A}_{i}$ are not differentiable with respect to $\pi_{x,y}$, where $\pi_{x,y}$ denotes the probability of an edge existing between nodes $x$ and $y$ in $\mathbf{A}_{u}$ or $\mathbf{A}_{i}$. Next, we employ the reparameterization trick \cite{2022vibgsl} to optimize the implicit graph structure $\mathbf{A}_u$ (or $\mathbf{A}_i$):
\begin{equation}
\begin{aligned}
    \mathbf{A}_u \text{ } \text{or} \text{ } \mathbf{A}_i &= \bigcup_{{x,y} \in \boldsymbol{V}_{u}} \{a_{x,y} \sim \text{Ber}(\pi_{x,y})\}, \\
    \text{Ber}(\pi_{x,y}) &\approx \text{sigmoid}\left( \frac{1}{t} \left( \log \frac{\pi_{x,y}}{1 - \pi_{x,y}} + \log \frac{\epsilon}{1 - \epsilon} \right) \right),
\end{aligned}
\label{eq10}
\end{equation}
where \(\epsilon\) follows a Uniform distribution $U(0,1)$ and \(t\) in \( \mathbb{R}^+ \) determines the temperature of the concrete distribution. We then implement a gradient-based autoencoder by setting thresholds $a_u$ and $a_i$ to filter the noise.

While directly computing mutual information (Eq. \ref{eq9}) is difficult, we complete this process by designing and optimizing the upper bound of $I(\mathbf{S}; \mathbf{A})$ and the lower bound of $(\mathbf{S}; \mathbf{Y}_{Rec})$.
\begin{itemize}[leftmargin=*]
\item \textbf{Upper bound.} To eliminate unnecessary noise, we maximize the mutual information between the encoder and decoder through $I(\mathbf{A}_u; \mathbf{S}_{u})$ and $I(\mathbf{A}_i; \mathbf{S}_{i})$. The upper bound of the autoencoder's embeddings can be represented by the following method:
\begin{equation}
\begin{aligned}
    I(\mathbf{A}_u; \mathbf{S}_{u}) &\leq \sum p(\mathbf{S}_{u}) p(\mathbf{A}_u|\mathbf{S}_{u}) \log \frac{p(\mathbf{A}_u|\mathbf{S}_{u})}{r(\mathbf{A}_u)}, \\
    I(\mathbf{A}_i; \mathbf{S}_{i}) &\leq \sum p(\mathbf{S}_{i}) p(\mathbf{A}_i|\mathbf{S}_{i}) \log \frac{p(\mathbf{A}_i|\mathbf{S}_{i})}{r(\mathbf{A}_i)},
\end{aligned}
\end{equation}
where $p(\cdot)$ is the probability distribution of the embeddings, and $p(A|B)$ denotes the conditional probability distribution. Moreover, $r(\cdot)$ is a Gaussian approximation of $p(\cdot)$.
\item \textbf{Lower bound.} The lower bound of the learned autoencoder's embeddings, based on the supervision of the downstream recommendation task $\mathbf{Y}_{Rec}$, can be expressed as:
\begin{equation}
\begin{aligned}
    I(\mathbf{S}_u; \mathbf{Y}_{Rec}) &\geq \sum p(\mathbf{Y}_{Rec}, \mathbf{S}_u) \log q(\mathbf{Y}_{Rec}|\mathbf{S}_u) + H(\mathbf{Y}_{Rec}), \\
    I(\mathbf{S}_i; \mathbf{Y}_{Rec}) &\geq \sum p(\mathbf{Y}_{Rec}, \mathbf{S}_i) \log q(\mathbf{Y}_{Rec}|\mathbf{S}_i) + H(\mathbf{Y}_{Rec}),
\end{aligned}
\end{equation}
where $H(\cdot)$ is a setting that is irrelevant for optimization. To approximate $p(A,B)$, we utilize the non-negativity \cite{2016dvib} of KL divergence to train $q(X|Y)$.
\end{itemize}

This is an example based on users to obtain the upper limit of $\mathcal{L}_{\text{UIB}}$, and similarly, we can obtain $\mathcal{L}_{\text{IIB}}$. To integrate Eq. \ref{eq9} along with its upper and lower bounds, we formulate the objective as minimizing the following part:
\begin{align*}
\mathcal{L}_{\text{UIB}} & =-I\left(\mathbf{S}_{u} ; \mathbf{Y}_{Rec} \right)+\beta \cdot I\left(\mathbf{S}_{u} ; \mathbf{A}_u \right) \\
& \leq-\sum \sum p\left(\mathbf{Y}_{Rec}, \mathbf{S}_{u} \right) \log q\left(\mathbf{Y}_{Rec} \mid \mathbf{S}_{u} \right) \\
& +\beta \cdot \sum \sum p\left(\mathbf{A}_u \right) p\left(\mathbf{S}_{u} \mid \mathbf{A}_u \right) \log \frac{p\left(\mathbf{S}_{u} \mid \mathbf{A}_u \right)}{r\left(\mathbf{S}_{u} \right)}, \tag{13}
\label{eq13}
\end{align*}

In order to approximate the upper limit mentioned above, we utilize the empirical distribution $p(\mathbf{Y}_{Rec}, \mathbf{A}_u)=p(\mathbf{A}_u) p(\mathbf{Y}_{Rec} \mid \mathbf{A}_u)=\frac{1}{\mathbf{N}} \sum_{n=1}^{\mathbf{N}} \delta \mathbf{Y}_{n}(\mathbf{Y}_{Rec}) \delta \mathbf{A}_{n}(\mathbf{A}_u)$, and $\mathbf{N}$ is sampling number. The $\mathcal{L}_{\mathrm{KL}}$ approximates the part mentioned in Eq. \ref{eq13}: $\mathcal{L}_{\text{UIB}} \leq \mathcal{L}_{\mathrm{KL}}$. The specific losses are as the follow:
\begin{align*}
    \mathcal{L}_{\text{UIB}} & = \frac{1}{\mathbf{N}} \sum_{n=1}^{\mathbf{N}} -\log q(\mathbf{Y}_{n} \mid \mathbf{S}_{u}) \\
    & + \beta \frac{1}{\mathbf{N}} \sum_{n=1}^{\mathbf{N}} p(\mathbf{S}_{u} \mid \mathbf{A}_{n}) \log \frac{p(\mathbf{S}_{u} \mid \mathbf{A}_{n})}{r(\mathbf{S}_{u})} , \tag{14}
    \label{eq14}
\end{align*}
where a distribution $p(\mathbf{S}_{u}\mid\mathbf{A}_{n})$ is $\mathcal{M}\left(\mathbf{S}_{u} \mid \mu\left(\mathbf{A}_{n}\right), \eta\left(\mathbf{A}_{n}\right)\right)$, and $\mathcal{M}$ is a Normal distribution. We leverage mean-pooing: 
\begin{equation}
    \left(\mu\left(\mathbf{A}_{n}\right), \eta\left(\mathbf{A}_{n}\right)\right)=\operatorname{Pooling}\left(\left\{\mathbf{S}_{u}^1, \mathbf{S}_{u}^2, \mathbf{S}_{u}\right\}\right), \tag{15}
    \label{eq15}
\end{equation}
where $\mu(\cdot)$ and $\eta(\cdot)$ are calculated mean and standard deviation. The Eq. \ref{eq15} represents the different user-based views, which $\mu\left(\mathbf{A}_{n}\right) \in \mathbb{R}^{M \times d / 2}$ and $\eta\left(\mathbf{A}_{n}\right) \in \mathbb{R}^{M \times d / 2}$. According to \cite{2016dvib, 2024graphaug}, this derivation process can be proven. We utilize the same derivation to obtain the  $\mathcal{L}_{\text{IIB}}$ based on items.

\subsection{\textbf{Dual Contrastive Learning (DCL)}}
In this section, we propose Dual Contrastive Learning (DCL) to enhance the existing graph contrastive learning frameworks, which lack multi-meta-path augmentation. DCL aims to align the diverse intents of users and items at the meta-path level. The view level integrates these intents into real interactions, capturing global consistency at a higher level.

Following empirical practices \cite{2021sgl, 2022simgcl}, we utilize the LightGCN for multi-layer iterations to obtain the user embeddings $\mathbf{E}_u^{'}$ and item embeddings $\mathbf{E}_i^{'}$, s detailed in Eqs. \ref{eq1} and \ref{eq2}. Next, we employ the obtained embeddings for DCL to achieve alignment between intent-intent contrast and intent-interaction contrast. Specifically, \textbf{User-item view:} $\mathbf{E}_u^{'}$ and $\mathbf{E}_i^{'}$; \textbf{Heterogeneous view:} $\mathbf{E}_{uu}^{1}$, $\mathbf{E}_{uu}^{2}$, $\mathbf{E}_{ii}^{1}$,$\mathbf{E}_{ii}^{2}$.
\begin{itemize}[leftmargin=*]
\item \textbf{Intent-intent contrast.}
The inputs for this part is the embeddings learned from the heterogeneous information view. We align the information of users and items respectively. Inspired by \cite{2023hgcml}, we adopt the InfoNCE \cite{2020simclr} loss as the alignment objective and maximize the mutual information between the two. For user $u$, we represent each node in $\mathbf{E}_{uu}^{1}$ as $\mathbf{e}^{'}$, and each node in $\mathbf{E}_{uu}^{2}$ as $\mathbf{e}^{''}$. The defined intent-intent contrast loss (ICL) is as follows:
\begin{equation}
    \mathcal{L}_{\text{ICL}}^{\text{user}} = \sum_{i \in \mathcal{B}} -\log \frac{\exp \left( s\left(\mathbf{e}^{'}_i, \mathbf{e}^{''}_i \right) / \tau^{'} \right)}{\sum_{j \in \mathcal{B}} \exp \left( s\left(\mathbf{e}^{'}_i, \mathbf{e}^{''}_j \right) / \tau^{'}\right)},
    \tag{16}
    \label{eq16}
\end{equation}
where $i$, $j$ are users/items in a sampled batch $\mathcal{B}$. $s(\cdot, \cdot)$ means the cosine similarity, and $\tau^{'}$ is a temperature coefficient. Following the same approach, we obtain the contrastive loss $\mathcal{L}_{\text{ICL}}^{\text{item}}$ of items. Meanwhile, some studies \cite{2022cgi, 2018rlcpd} indicates that minimizing the InfoNCE loss corresponds to increasing the mutual information. We utilize positive and negative samples obtained from sampled batch $\mathcal{B}$ to calculate the mutual information between different meta-paths on the user side (or item side). Intent-intent contrast aims to align specific intents (meta-paths) and capture similarities between different intents.
\item \textbf{Intent-interaction contrast.}
The inputs for this part consists of embeddings learned from combining user-item and heterogeneous views, denoted as: $\left(\mathbf{E}_{u}^{'} + \mathbf{E}_{uu}^{1}\right)$ and $\left(\mathbf{E}_{u}^{'} + \mathbf{E}_{uu}^{2}\right)$. For user $u$, we denote each nodes by $\mathbf{z}^{'}$ and $\mathbf{z}^{''}$. Considering the user preferences and the item relationships, we model these intents using heterogeneous information. Intent refers to the motivation behind a user's choice of items, and this process is often traceable. For example, two users who happen to belong to the same age group and occupation are empirically more likely to interact with similar items. To learn the intent representations behind the views enhanced by multiple meta-paths, we use the same loss function to further align specific users (or items):
\begin{equation}
    \mathcal{L}_{\text{IICL}}^{\text{user}} = \sum_{i \in \mathcal{B}} -\log \frac{\exp \left( s\left(\mathbf{z}^{'}_i, \mathbf{z}^{''}_i \right) / \tau^{''} \right)}{\sum_{j \in \mathcal{B}} \exp \left( s\left(\mathbf{z}^{'}_i, \mathbf{z}^{''}_j \right) / \tau^{''}\right)},
    \tag{17}
    \label{eq17}
\end{equation}
where $\tau^{''}$ is the temperature coefficient for intent-interaction contrast. Based on the similarly enhanced embeddings $\left(\mathbf{E}_{i}^{'} + \mathbf{E}_{ii}^{1}\right)$ and $\left(\mathbf{E}_{i}^{'} + \mathbf{E}_{ii}^{2}\right)$, we can obtain the loss $\mathcal{L}_{\text{IICL}}^{\text{item}}$ for items.
\end{itemize}

\subsection{\textbf{Model Optimization}}
To improve self-supervised recommendation, we utilize a multi-task joint learning to formulate the final optimization objective. First, we organize the loss functions of the main modules: BAE and DCL. BAE employs the information bottleneck principle to denoise in the autoencoder, from which we can obtain:
\begin{align*}
    \mathcal{L}_{\text{IB}} = \mathcal{L}_{\text{UIB}} + \mathcal{L}_{\text{IIB}},
    \tag{18}
\end{align*}

Since the users' and items' views are symmetrical, we do not set separate coefficients for them. However, DCL has a hierarchical structure, which we summarize as follows:
\begin{align*}
    \mathcal{L}_{\text{DCL}} &= \lambda_{\text{ICL}} \cdot \mathcal{L}_{\text{ICL}} + \lambda_{\text{IICL}} \cdot \mathcal{L}_{\text{IICL}} \\
    &= \lambda_{\text{ICL}} \cdot \left(\mathcal{L}_{\text{ICL}}^{\text{user}} + \mathcal{L}_{\text{ICL}}^{\text{item}} \right) + \lambda_{\text{IICL}} \cdot \left(\mathcal{L}_{\text{IICL}}^{\text{user}} + \mathcal{L}_{\text{IICL}}^{\text{item}} \right),
    \tag{19}
\end{align*}
where $\lambda_{\text{ICL}}$ and $\lambda_{\text{IICL}}$ represent the weights for intent-intent contrast and intent-interaction contrast, respectively. Next, after summarizing the loss functions of the modules we proposed, we introduce the pairwise Bayesian Personalized Ranking (BPR) loss \cite{2012bpr} commonly used in recommendation task:
\begin{equation}
    \mathcal{L}_{\text{BPR}}=-\frac{1}{|\mathcal{B}|} \sum_{(i, j, k) \in B} \log \sigma\left(\mathbf{d}_{i}^{\top} \mathbf{d}_{j}-\mathbf{d}_{i}^{\top} \mathbf{d}_{k}\right),
    \tag{20}
    \label{eq20}
\end{equation}
where $\mathcal{B}=\left\{(i, j, k) \mid A_{i, j}=1, A_{i, k}=0\right\}$ is the training data, and embeddings $\mathbf{d} \in \left\{\mathbf{E}_u^{'}+\mathbf{E}_{uu}^{1}+\mathbf{E}_{uu}^{2}\right\}$ are obtained by weighting the main task and heterogeneous information. This loss function enforces that the predicted scores for observed interactions are higher than those for unobserved interactions. Finally, the complete optimization objective of IHGCL is as the follow:
\begin{equation}
    \mathcal{L}_{\text{IHGCL}} = \mathcal{L}_{\text{BPR}} + \lambda_1 \cdot \mathcal{L}_{\text{IB}} + \mathcal{L}_{\text{DCL}} + \lambda_2 \cdot \|\Theta\|_{2}^{2} 
    \tag{21}
    \label{eq21}
\end{equation}
where $\lambda_1$ and $\lambda_2$ are adjustable weights, and $\|\Theta\|_{2}^{2}$ are trainable model parameters and $L_2$ regularization. Because $\lambda_{\text{ICL}}$ and $\lambda_{\text{IICL}}$ have already determined the weight of $\mathcal{L}_{\text{DCL}}$, there is no need to set them again here. Algorithm \ref{algorithm1} provides the overall process of the learning steps for IHGCL.

\begin{algorithm}[ht]
\caption{The IHGCL Learning Algorithm}
\begin{algorithmic}[1]
\item[] \textbf{Input:} User-item interaction matrix $\mathbf{A}$, meta-path-based subgraph $\mathbf{A}_u^1$, $\mathbf{A}_u^2$, $\mathbf{A}_i^1$, $\mathbf{A}_i^2$, maximum training epochs $E$ and parameters required for the training process
\item[] \textbf{Output:} Trained node embeddings
\STATE Initialize all parameters;
\FOR{\text{epoch} in \{1, 2, \ldots, $E$\}}
    \STATE Obtain high-order embedding $\mathbf{E}_u^{'}$ and $\mathbf{E}_i^{'}$ of the input graph $\mathbf{A}$ via $l$-layer LightGCN, applying Eq. \ref{eq2};
    \STATE Obtain intent embeddings $\mathbf{E}_{uu}^{1}$, $\mathbf{E}_{uu}^{2}$, $\mathbf{E}_{ii}^{1}$, $\mathbf{E}_{ii}^{2}$ using the \textbf{BAE} module via meta-path-based subgraph $\mathbf{A}_u^1$, $\mathbf{A}_u^2$, $\mathbf{A}_i^1$, $\mathbf{A}_i^2$, applying Eqs. \ref{eq5}--\ref{eq8};
    \STATE Learn distributions of $\mu(\mathbf{A}_n), \eta(\mathbf{A}_n)$ via pooling operation on $\mathbf{S}_{u}^1$, $\mathbf{S}_{u}^2$ via Eq. \ref{eq15};
    \STATE Optimize \textbf{IB} via loss $\mathcal{L}_{\text{IB}}$ according to Eq. \ref{eq14};
    \STATE Perform \textbf{DCL} augmentation based on two views via loss $\mathcal{L}_{\text{ICL}}$ and $\mathcal{L}_{\text{IICL}}$ according to Eqs. \ref{eq16} and \ref{eq17};
    \STATE Optimize BPR loss $\mathcal{L}_{\text{BPR}}$ via parameter regularizer via Eq. \ref{eq20};
    \STATE Joint optimization of IHGCL following Eq. \ref{eq21};
\ENDFOR
\RETURN all parameters and user and item embeddings;
\end{algorithmic}
\label{algorithm1}
\end{algorithm}

\section{Model Analysis}
\subsection{\textbf{Theoretical Analysis.}}
In this section, we perform an in-depth analysis of IHGCL to answer how recommendation tasks can benefit from heterogeneous information. We decompose the model into the two proposed components: DCL and BAE. 

\textbf{DCL:}
We utilize heterogeneous information to model the multi-view intent of users and items for the main task augmentation. This process is formally described as follows: Given a user node $u$ and its initialized representation $\mathbf{e}_u$ in the $d$-dimensional embedding space (Eq. \ref{eq17}):
\begin{align*}
    \mathbf{z}'_u &= \mathbf{e}_u^{'} + \mathbf{e}_{uu}^1 = \mathcal{F}_{ui}\left(\mathbf{e}_u\right) + \mathcal{F}_{uu}^{'}\left(\mathbf{e}_u\right), \\
    \mathbf{z}''_u &= \mathbf{e}_u^{'} + \mathbf{e}_{uu}^2 = \mathcal{F}_{ui}\left(\mathbf{e}_u\right) + \mathcal{F}_{uu}^{''}\left(\mathbf{e}_u\right),
    \tag{22}
    \label{eq22}
\end{align*}
where $\mathcal{F}_{uu}^{'}$ and $\mathcal{F}_{uu}^{''}$ are BAE, and $\mathcal{F}_{ui}$ is LightGCN encoder. 

\begin{figure}[t]
    \centering
    \begin{minipage}[b]{0.48\linewidth}
        \centering
        \subfloat[\small DCL]{\includegraphics[width=\linewidth]{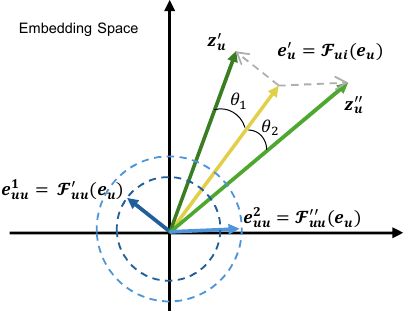}}
    \end{minipage}
    \begin{minipage}[b]{0.48\linewidth}
        \centering
        \subfloat[\small BAE]{\includegraphics[width=\linewidth]{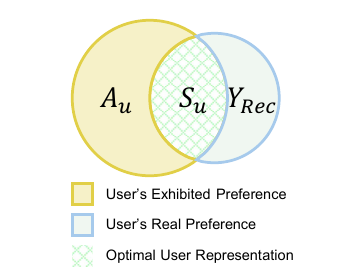}}
    \end{minipage}
    \caption{Theoretical analysis of DCL and BAE.}
    \label{fig5:analysis}
    \vspace{-0.5cm}
\end{figure}

As illustrated in Figure \ref{fig5:analysis} (a), we assume that in the contrastive loss, $\mathcal{F}_{ui}$ encodes the multi-hop information of user-item interactions, while $\mathcal{F}_{uu}$ models the preferences of users contained within different meta-paths in heterogeneous information. These preferences (blue) interact with embeddings in the interaction domain (yellow), leading in corresponding deviations ($\theta_1$ and $\theta_2$) to produce enhanced embeddings. Notably, the process of contrastive learning, which aligns and integrates embeddings with various intents, aids the model in better perceiving the intents of users and items. This point will be validated in Section \ref{3.3}. Therefore, having high-quality user and item intents is crucial for effective contrastive learning, which explains the necessity of introducing BAE.

\textbf{BAE:}
We combine the techniques of masked autoencoder and information bottleneck (IB) for denoising, aiming to learn embeddings of users or items devoid of irrelevant information. The mutual information in Figure \ref{fig5:analysis} (b) can be expressed as:
\begin{equation}
     \mathcal{L}_{\text{IB}}(\mathbf{S}_u, \mathbf{A}_u; \mathbf{Y}_{Rec}) = - I(\mathbf{S}_{u}; \mathbf{Y}_{Rec}) + \beta \cdot I(\mathbf{A}_u; \mathbf{S}_{u}), \tag{23}
\end{equation}
where $\mathbf{S}_u$ denotes the weighted embeddings of the encoder and decoder, which are the user preferences we need to optimize. The connections brought by the meta-paths merely represent the users' exhibited preferences ($\mathbf{A}_u$), which often contain noise, while $\mathbf{Y}_{Rec}$ is the downstream recommendation information. The IB strategy encourages the representation to capture the minimal sufficient information needed for downstream tasks, which involves maximizing the intersection between $\mathbf{S}_u$ and $\mathbf{Y}_{Rec}$, and minimizing the intersection between $\mathbf{S}_u$ and $\mathbf{A}_u$. Specifically, Eqs. \ref{eq9} to \ref{eq15} illustrate the denoised process for heterogeneous information.

\subsection{\textbf{Time Complexity Analysis.}}
This section discusses the batch time complexity \cite{2022simgcl}. The user-item view's time complexity is $\mathcal{O} \left(2|\mathcal{V}_{ui}| \cdot L d\right)$, where $|\mathcal{V}_{ui}|$ denotes the edge number of a interaction graph $\mathbf{R}$. $L$ is the GCN layers and $d$ is the feature dimension. For the heterogeneous view, IHGCL has four meta-path-based subgraphs: $G_U^{1}, G_U^{2}, G_I^{1}, G_I^{2}$ (Eq. \ref{eq5}). So, the time complexity of the heterogeneous view is $\mathcal{O}\left(2|\boldsymbol{V}_u| \cdot L^{'} d + 2|\boldsymbol{V}_i| \cdot L^{'} d\right)$, where $L{'}$ is the number of autoencoder layers $\left(L_E+L_D\right)$. Then, the self-supervised loss contains $\lambda_{\text{IB}}$ and $\lambda_{\text{DCL}}$. The time complexity of $\lambda_{\text{IB}}$ primarily depends on the number of nodes and can generally be considered as $\mathcal{O}\left(Md+Nd\right)$ in most cases. To calculate the complexity of $\lambda_{\text{DCL}}$, we get doubled time $\mathcal{O}\left(2Bd+2BCd\right)$ compared to general contrastive learning, where $C$ represents the node number in a batch and $B$ denote the batch size. In the recommendation task, the complexity of the additional $\mathcal{L}_{\text{BPR}}$ is $\mathcal{O}\left(2Bd\right)$. Thus, the overall time complexity of encoding is $\mathcal{O}\left(2(|\mathcal{V}_{ui}| \cdot L + |\boldsymbol{V}_u| \cdot L^{'} + |\boldsymbol{V}_i| \cdot L^{'} )d\right)$, and losses complexity are $\mathcal{O}\left((M+N+4B+2BC)d\right)$. This complexity is the time complexity for each epoch during training, and the number of training epochs determine the overall training time.

\begin{table}[t]
    \captionsetup{justification=centering}
    \caption{Statistics of datasets used in this paper.}
    \begin{adjustbox}{width=0.5\textwidth}
    \begin{NiceTabular}{cccccc}
    \midrule Dataset & User \# & Item \# & Interaction \# & Meta-paths \\
    \midrule 
    Last.fm & 1,892 & 17,632 & 92,834 & UU, UATAU, AA, ATA \\
    Amazon & 6,170 & 2,753 & 195,791 & UIU, UIBIU, IBI, ICI \\
    Yelp & 16,239 & 14,284 & 198,397 & UU, UBU, BCiB, BCaB \\
    Douban Book & 13,024 & 22,347 & 792,062 & UU, UGU, BAB, BYB \\
    Movielens-1M & 6040 & 3952 & 1,000,209 & UMU, UOU, MUM, MGM \\
    Douban Movie & 13,367 & 12,677 & 1,068,278 & UU, UGU, MAM, MDM \\
    \midrule
    \end{NiceTabular}
    \end{adjustbox}
    \label{table1:datasets}
    \vspace{-0.5cm}
\end{table}

\begin{table*}[t]
    \captionsetup{justification=centering}
    \caption{Performance Comparison of Different Recommendation Methods with Varying Top-N on Six Datasets.\\
    \small The best results are shown in bold and the second-best results are underlined. R@K represents Recall@K, and N@K represents NDCG@K, and the improvement is significant based on two-tailed paired t-test.}
    \begin{adjustbox}{width=\textwidth}
    \begin{NiceTabular}{cc|cccc|ccc|cccccccc}
    \toprule[1pt]
    \multirow{2}{*}{ Datasets} & \multirow{2}{*}{ Metrics} & \multicolumn{4}{c}{Graph Recommendation Models} & \multicolumn{3}{c}{Intent Recommendation Models} & \multicolumn{7}{c}{Heterogeneous Graph Recommendation Models} \\
    \cmidrule{3-16} & & LightGCN & SGL & SimGCL  & LightGCL & DisenHAN & DCCF & BIGCF & HERec & HAN & HeCo & SMIN & HGCL & IHGCL & $p$-value \\
    \hline \midrule \multirow{6}{*}{ Last.fm }
    & R@5  & 0.1242 & 0.1293 & 0.1335 & 0.1302 & 0.1158 & 0.1306 & $\underline{0.1347}$ & 0.1117 & 0.1084 & 0.1169 & 0.1231 & $0.1298$& $\mathbf{0.1370}$ & $4.7e^{-5}$ \\
    & N@5  & 0.2673 & 0.2786 & $\underline{0.2859}$ & 0.2805 & 0.2581 & 0.2804 & 0.2852 & 0.2386 & 0.2291 & 0.2565 & 0.2671 & $0.2777$& $\mathbf{0.2913}$ & $9.2e^{-4}$ \\
    & R@10 & 0.1866 & 0.1923 & $\underline{0.1966}$ & 0.1938 & 0.1772 & 0.1945 & 0.1959 & 0.1704 & 0.1678 & 0.1757 & 0.1818 & $0.1907$& $\mathbf{0.2044}$ & $3.3e^{-3}$ \\
    & N@10 & 0.2356 & 0.2404 & 0.2467 & 0.2419 & 0.2234 & 0.2429 & $\underline{0.2476}$ & 0.2093 & 0.2025 & 0.2226 & 0.2299 & $0.2390$& $\mathbf{0.2542}$ & $6.5e^{-6}$ \\
    & R@20 & 0.2626 & 0.2714 & $\underline{0.2760}$ & 0.2730 & 0.2516 & 0.2734 & 0.2752 & 0.2468 & 0.2382 & 0.2514 & 0.2581 & $0.2698$& $\mathbf{0.2824}$ & $8.8e^{-4}$ \\
    & N@20 & 0.2628 & 0.2694 & 0.2755 & 0.2711 & 0.2492 & 0.2722 & $\underline{0.2764}$ & 0.2385 & 0.2284 & 0.2497 & 0.2574 & $0.2683$& $\mathbf{0.2815}$ & $2.4e^{-8}$ \\
    \midrule \multirow{6}{*}{ Amazon } 
    & R@5  & 0.0653 & 0.0704 & 0.0742 & 0.0709 & 0.0608 & 0.0718 & $\underline{0.0743}$ & 0.0571 & 0.0546 & 0.0618 & 0.0640 & $0.0706$& $\mathbf{0.0762}$ & $5.2e^{-7}$ \\
    & N@5  & 0.0875 & 0.0948 & 0.1011 & 0.0974 & 0.0821 & 0.0990 & $\underline{0.1014}$ & 0.0766 & 0.0748 & 0.0840 & 0.0873 & $0.0978$& $\mathbf{0.1040}$ & $5.8e^{-4}$ \\
    & R@10 & 0.1028 & 0.1084 & $\underline{0.1197}$ & 0.1038 & 0.0958 & 0.1147 & 0.1193 & 0.0903 & 0.0885 & 0.0995 & 0.1031 & $0.1094$& $\mathbf{0.1230}$ & $3.2e^{-6}$ \\
    & N@10 & 0.0976 & 0.1041 & $\underline{0.1138}$ & 0.0987 & 0.0905 & 0.1061 & 0.1127 & 0.0853 & 0.0832 & 0.0934 & 0.0969 & $0.1062$& $\mathbf{0.1161}$ & $7.8e^{-6}$ \\
    & R@20 & 0.1592 & 0.1646 & 0.1751 & 0.1574 & 0.1508 & 0.1705 & $\underline{0.1758}$ & 0.1420 & 0.1385 & 0.1519 & 0.1569 & $0.1670$& $\mathbf{0.1805}$ & $8.2e^{-8}$ \\
    & N@20 & 0.1151 & 0.1215 & 0.1306 & 0.1140 & 0.1098 & 0.1255 & $\underline{0.1307}$ & 0.1017 & 0.0988 & 0.1097 & 0.1135 & $0.1247$& $\mathbf{0.1346}$ & $5.3e^{-5}$ \\
    \midrule \multirow{6}{*}{ Yelp }
    & R@5  & 0.0350 & 0.0371 & 0.0398 & 0.0373 & 0.0317 & 0.0379 & $\underline{0.0405}$ & 0.0282 & 0.0264 & 0.0318 & 0.0336 & $0.0367$& $\mathbf{0.0425}$ & $4.7e^{-6}$ \\
    & N@5  & 0.0406 & 0.0414 & 0.0441 & 0.0421 & 0.0349 & 0.0427 & $\underline{0.0446}$ & 0.0330 & 0.0316 & 0.0352 & 0.0352 & $0.0400$& $\mathbf{0.0473}$ & $6.4e^{-5}$ \\
    & R@10 & 0.0584 & 0.0622 & $\underline{0.0661}$ & 0.0631 & 0.0519 & 0.0632 & 0.0658 & 0.0460 & 0.0423 & 0.0523 & 0.0540 & $0.0628$& $\mathbf{0.0693}$ & $2.8e^{-4}$ \\
    & N@10 & 0.0471 & 0.0509 & $\underline{0.0525}$ & 0.0512 & 0.0409 & 0.0504 & 0.0519 & 0.0378 & 0.0357 & 0.0412 & 0.0409 & $0.0512$& $\mathbf{0.0546}$ & $4.0e^{-5}$ \\
    & R@20 & 0.0883 & 0.0961 & 0.1003 & 0.0980 & 0.0820 & 0.0973 & $\underline{0.1009}$ & 0.0766 & 0.0734 & 0.0823 & 0.0854 & $0.0963$& $\mathbf{0.1044}$ & $5.1e^{-7}$ \\
    & N@20 & 0.0555 & 0.0604 & 0.0616 & 0.0608 & 0.0500 & 0.0605 & $\underline{0.0620}$ & 0.0469 & 0.0440 & 0.0497 & 0.0500 & $0.0605$& $\mathbf{0.0649}$ & $2.9e^{-6}$ \\
    \midrule
    & R@5  & 0.0626 & 0.0728 & $\underline{0.0752}$ & 0.0726 & 0.0567 & 0.0725 & 0.0742 & 0.0509 & 0.0480 & 0.0564 & 0.0597 & $0.0708$& $\mathbf{0.0800}$ & $6.3e^{-6}$ \\
    & N@5  & 0.1353 & 0.1547 & $\underline{0.1560}$ & 0.1534 & 0.1274 & 0.1541 & 0.1551 & 0.1027 & 0.0984 & 0.1284 & 0.1330 & $0.1509$& $\mathbf{0.1616}$ & $7.7e^{-5}$ \\
    \multirow{1}{*}{ Douban }
    & R@10 & 0.0968 & 0.1136 & 0.1147 & 0.1130 & 0.0909 & 0.1137 & $\underline{0.1149}$  & 0.0836 & 0.0801 & 0.0905 & 0.0934 & $0.1103$& $\mathbf{0.1190}$ & $5.8e^{-4}$ \\
    \multirow{1}{*}{ Book }
    & N@10 & 0.1312 & 0.1507 & 0.1519 & 0.1495 & 0.1245 & 0.1506 & $\underline{0.1535}$ & 0.1042 & 0.0975 & 0.1248 & 0.1298 & $0.1470$& $\mathbf{0.1556}$ & $5.1e^{-4}$ \\
    & R@20 & 0.1471 & 0.1665 & $\underline{0.1701}$ & 0.1666 & 0.1384 & 0.1665 & 0.1679 & 0.1312 & 0.1255 & 0.1387 & 0.1433 & $0.1627$& $\mathbf{0.1726}$ & $3.2e^{-7}$ \\
    & N@20 & 0.1370 & 0.1551 & 0.1579 & 0.1556 & 0.1311 & 0.1556 & $\underline{0.1591}$ & 0.1130 & 0.1076 & 0.1299 & 0.1339 & $0.1547$& $\mathbf{0.1611}$ & $3.8e^{-5}$ \\
    \midrule \multirow{6}{*}{ Movielens } 
    & R@5  & 0.0960 & 0.0992 & 0.1042 & 0.1002 & 0.0912 & 0.1006 & $\underline{0.1048}$ & 0.0871 & 0.0813 & 0.0912 & 0.0930 & $0.0985$& $\mathbf{0.1071}$ & $2.5e^{-6}$ \\
    & N@5  & 0.4180 & 0.4235 &0.4366 & 0.4281 & 0.4102 & 0.4307 &  $\underline{0.4376}$ & 0.3873 & 0.3865 & 0.4124 & 0.4150 & $0.4256$& $\mathbf{0.4482}$ & $4.4e^{-5}$ \\
    & R@10 & 0.1593 & 0.1648 & $\underline{0.1721}$ & 0.1654 & 0.1505 & 0.1667 & 0.1680 & 0.1444 & 0.1331 & 0.1516 & 0.1521 & $0.1633$& $\mathbf{0.1764}$ & $6.0e^{-4}$ \\
    & N@10 & 0.3903 & 0.3999 & $\underline{0.4087}$ & 0.4035 & 0.3811 & 0.4034 & 0.4050 & 0.3613 & 0.3550 & 0.3824 & 0.3828 & $0.4075$& $\mathbf{0.4172}$ & $5.3e^{-7}$ \\
    & R@20 & 0.2509 & 0.2569 & $\underline{0.2667}$ & 0.2578 & 0.2355 & 0.2606 & 0.2625 & 0.2303 & 0.2127 & 0.2374 & 0.2378 & $0.2530$& $\mathbf{0.2713}$ & $6.2e^{-6}$ \\
    & N@20 & 0.3763 & 0.3837 & $\underline{0.3942}$ & 0.3863 & 0.3649 & 0.3869 & 0.3916 & 0.3494 & 0.3376 & 0.3664 & 0.3666 & $0.3843$& $\mathbf{0.4010}$ & $8.1e^{-7}$ \\
    \midrule 
    & R@5  & 0.0700 & 0.0764 & $\underline{0.0818}$ & 0.0774 & 0.0693 & 0.0786 & 0.0816 & 0.0683 & 0.0663 & 0.0692 & 0.0695 & $0.0752$& $\mathbf{0.0847}$ & $4.0e^{-3}$ \\
    & N@5  & 0.2085 & 0.2098 & $\underline{0.2174}$ & 0.2111 & 0.2050 & 0.2103 & 0.2153 & 0.1778 & 0.1710 & 0.2064 & 0.1895 & $0.2099$& $\mathbf{0.2265}$ & $9.5e^{-5}$ \\
    \multirow{1}{*}{ Douban }
    & R@10 & 0.1139 & 0.1202 & $\underline{0.1258}$ & 0.1220 & 0.1114 & 0.1223 & 0.1253 & 0.1102 & 0.1018 & 0.1110 & 0.1123 & $0.1180$& $\mathbf{0.1274}$ & $5.5e^{-4}$ \\
    \multirow{1}{*}{ Movie }
    & N@10 & 0.2007 & 0.2025 & 0.2093 & 0.2036 & 0.1981 & 0.2051 & $\underline{0.2108}$ & 0.1764 & 0.1719 & 0.1974 & 0.1838 & $0.2021$& $\mathbf{0.2253}$ & $4.6e^{-5}$ \\
    & R@20 & 0.1781 & 0.1880 & $\underline{0.1972}$ & 0.1893 & 0.1720 & 0.1904 & 0.1963 & 0.1757 & 0.1692 & 0.1736 & 0.1743 & $0.1849$& $\mathbf{0.2003}$ & $1.8e^{-7}$ \\
    & N@20 & 0.2001 & 0.2046 & $\underline{0.2158}$ & 0.2052 & 0.1978 & 0.2084 & 0.2144 & 0.1828 & 0.1774 & 0.1961 & 0.1860 & $0.2037$& $\mathbf{0.2195}$ & $9.5e^{-5}$ \\
    \hline \bottomrule[1pt]
    \end{NiceTabular}
    \end{adjustbox}
    \label{table2:results}
\end{table*}

\section{EXPERIMENT}
In this section, we conduct extensive experiments and answer the following research questions:
\begin{itemize}[leftmargin=*]
\item \textbf{RQ1:} How does IHGCL compare to the current state-of-the-art (SOTA) models in terms of recommendation performance?
\item \textbf{RQ2:} What are the reasons for IHGCL's superior performance, and how does it differ from existing models?
\item \textbf{RQ3:} What impact does the selection of meta-paths have on the model?
\item \textbf{RQ4:} Are the key components in our IHGCL delivering the expected performance gains?
\item \textbf{RQ5:} How do different hyperparameter settings affect IHGCL?
\end{itemize}

\subsection{Experiment Setup}
\subsubsection{Datasets}
The performance of IHGCL has been validated on six real-world datasets. A statistical overview of all the datasets is presented in Table \ref{table1:datasets}.
We utilize various datasets for recommender systems: \textbf{Last.fm}\footnote{\url{https://www.last.fm/}} for music, \textbf{Amazon} \cite{2018herec} for products, \textbf{Yelp} \cite{2023chest} for businesses, \textbf{Douban} for books \cite{2018herec} and movies\footnote{\url{https://m.douban.com/}} and \textbf{Movielens-1M}\footnote{\url{https://grouplens.org/datasets/movielens/}} for movies, predicting user interactions with diverse entities like artists, items, businesses, books and movies.

\subsubsection{Baselines}
We respectively select the most representative baseline model for comparison.
\begin{itemize}[leftmargin=*]
\item \textbf{Graph Recommendation Models:} LightGCN \cite{2020lightgcn}, SGL \cite{2021sgl}, SimGCL \cite{2022simgcl} and LightGCL \cite{2023lightgcl}.
\item \textbf{Intent Recommendation Models:} DisenHAN \cite{2020disenhan}, DCCF \cite{2023dccf} and BIGCF \cite{2024bigcf}.
\item \textbf{Heterogeneous Graph Recommendation Models:} HERec \cite{2018herec}, HAN \cite{2019han}, HeCo \cite{2021heco}, SMIN \cite{2021smin} and HGCL \cite{2023hgcl}.
\end{itemize}

\subsubsection{Implementations}
To ensure a fair comparison, we follow the sampling method and dataset format of classic works \cite{2021sgl, 2020lightgcn, 2024dvgrl}. All models (including baselines) are retrained until convergence using the Adam optimizer and Xavier \cite{2010xavier} initialization for embeddings. For some classic heterogeneous graph neural network models, such as HAN \cite{2019han} and HeCo \cite{2021heco}, we replace the supervised loss with the BPR \cite{2012bpr} loss. We tune the parameters for the baselines on each dataset to ensure they achieve optimal performance. For general settings, GCN-based models maintain consistent layer numbers, learning rate 0.001, embedding size 64, and fixed batch size 4096. For the unique parameters of IHGCL, we analyze them in RQ4.


\subsection{Overall Performance (RQ1)}
Table \ref{table2:results} reports the recommendation performance of all baseline models on six public datasets, and we summarize possible observations or explanations for these outcomes. 
\begin{itemize}[leftmargin=*]
\item IHGCL demonstrates the best performance across all metrics on all datasets. Quantitatively, compared to the current SOTA baselines of HG-based recommendation, we achieved a significant improvement (douban movie for 11.48\% and amazon for 12.43\%). These experimental results prove the superiority and rationality of the proposed IHGCL.
\item IHGCL achieves substantial advancements over the models based on HGNN (e.g., HGCL \cite{2023hgcl} and SMIN \cite{2021smin}). This progress demonstrates the necessity of constructing multiple views of users and items through heterogeneous information for contrast.
\item IHGCL exhibits a distinct performance advantage over general recommender systems (e.g., SGL \cite{2021sgl} and SimGCL \cite{2022simgcl}). We have overcome the significant noise issues inherent in heterogeneous information and effectively utilized this information to extract user and item intents. The experimental results prove that IHGCL can better mine user preferences and item attributes for enhancing CL-based recommendation.
\item Considering the effectiveness in intent modeling, we compared the latest works in heterogeneous graph and bipartite graph scenarios. It can be observed that BIGCF \cite{2024bigcf} achieved better results than SimGCL on multiple metrics, proving that personalized intent recommendations can help users find what they truly need. Meanwhile, IHGCL uses meta-paths as intent-guided links to connect users or items, enhancing the model's ability to capture intents.
\end{itemize}

\begin{figure*}[t]
    \centering
    \begin{minipage}[b]{0.24\linewidth}
        \centering
        \subfloat[\small Amazon]{\includegraphics[width=\linewidth]{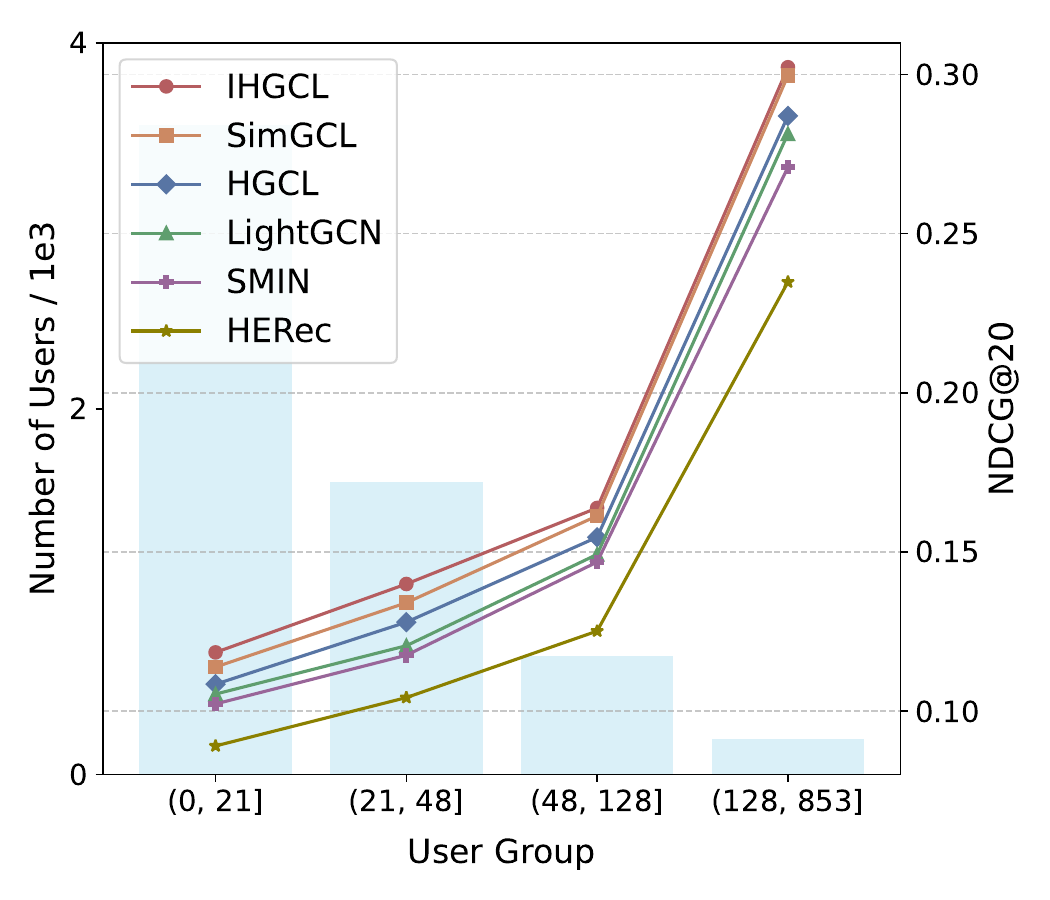}}
    \end{minipage}
    \begin{minipage}[b]{0.24\linewidth}
        \centering
        \subfloat[\small Yelp]{\includegraphics[width=\linewidth]{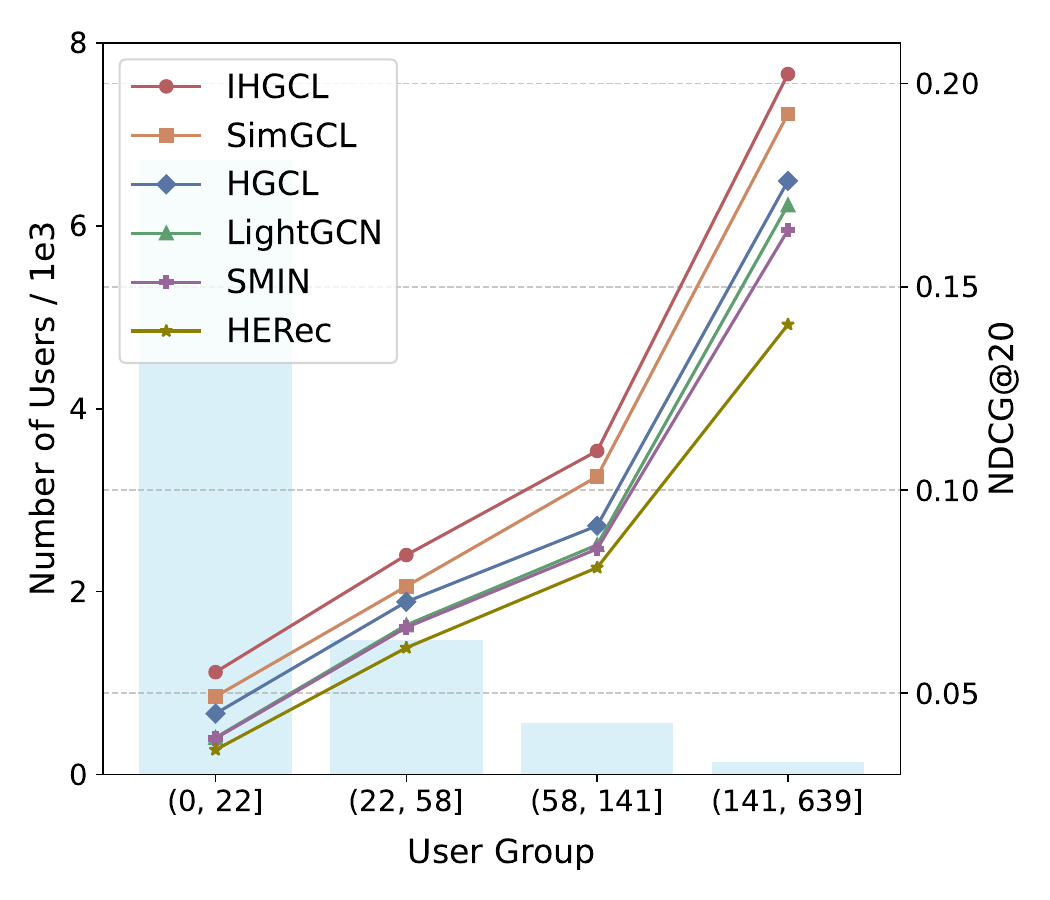}}
    \end{minipage}
    \begin{minipage}[b]{0.24\linewidth}
        \centering
        \subfloat[\small Douban Book]{\includegraphics[width=\linewidth]{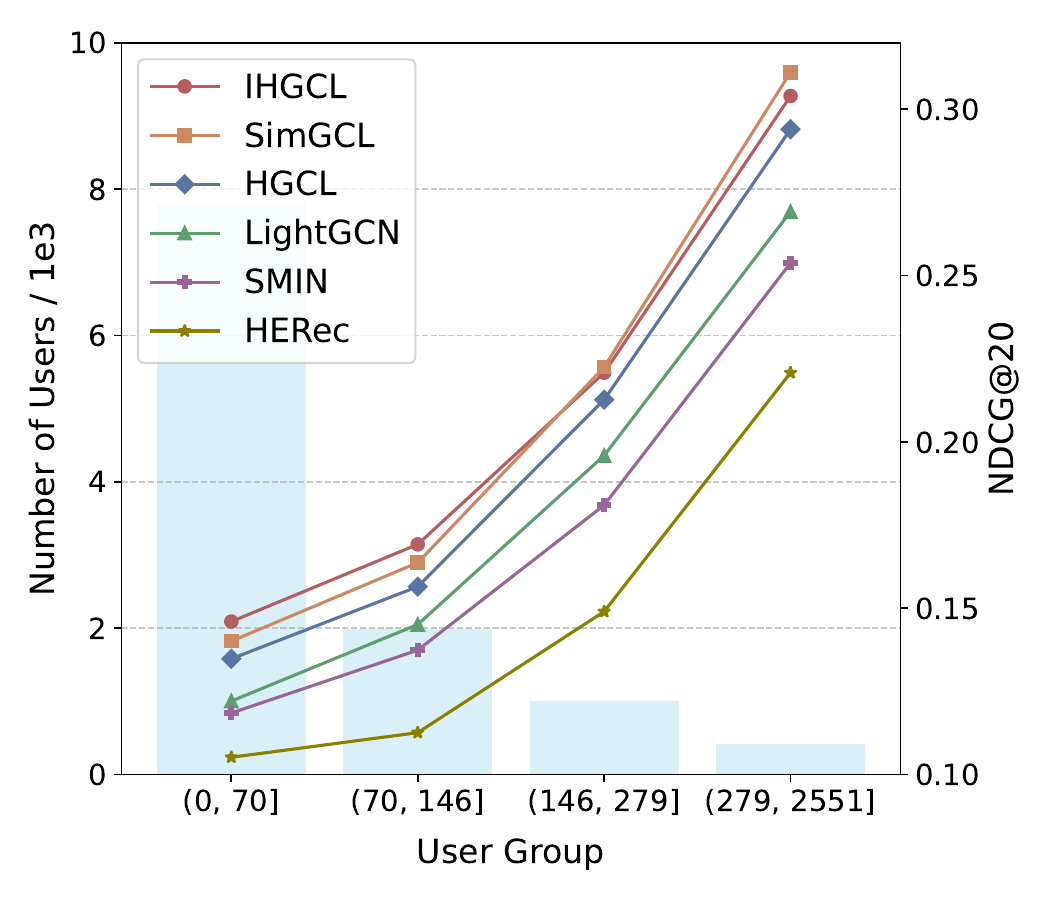}}
    \end{minipage}
    \begin{minipage}[b]{0.24\linewidth}
        \centering
        \subfloat[\small Douban Movie]{\includegraphics[width=\linewidth]{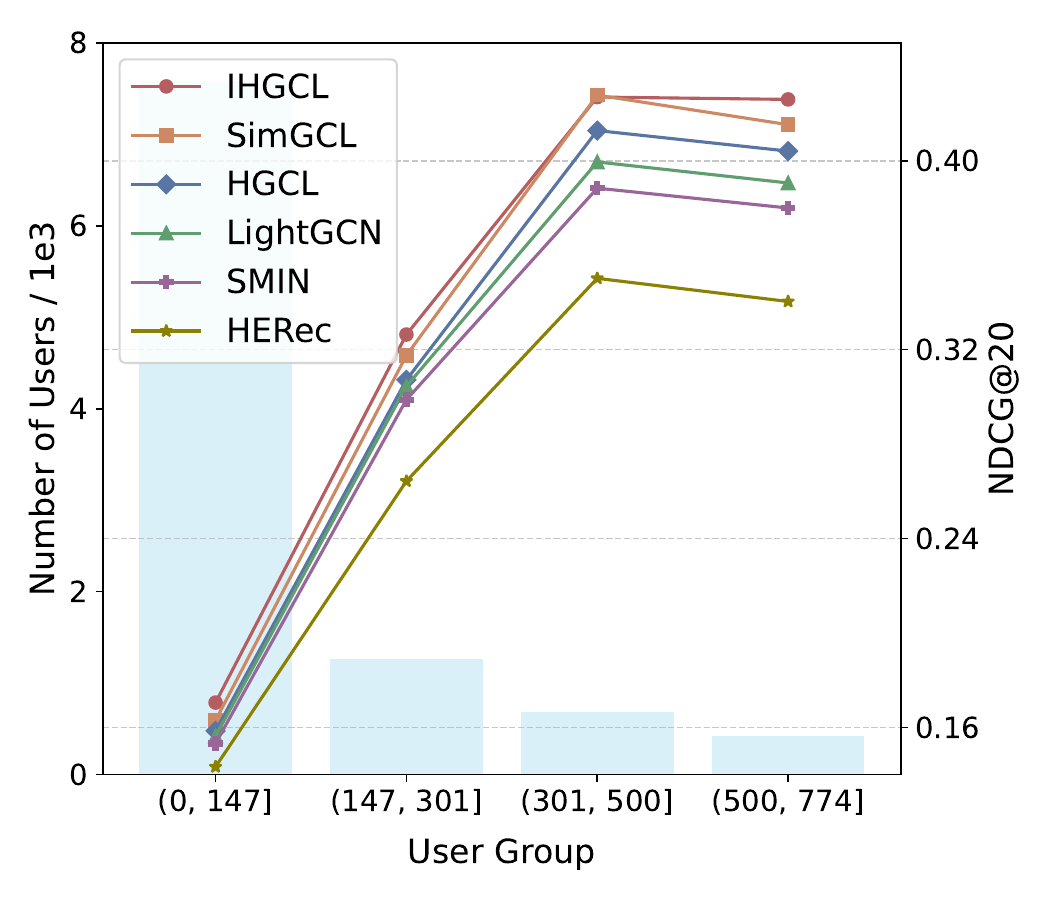}}
    \end{minipage}
    \caption{Performance comparison of different sparsity levels. The bar graph shows users' number per group on the left y-axis, and the line graph shows the performance of each method w.r.t. NDCG@20 on the right y-axis, with the x-axis denoting the interval of interactions per user group.}
    \label{fig6:sparsity}
    \vspace{-0.5cm}
\end{figure*}

\begin{figure}[t]
    \centering
    \begin{minipage}[b]{\linewidth}
        \centering
        \subfloat[\small Distribution of representations on Yelp]{\includegraphics[width=\linewidth]{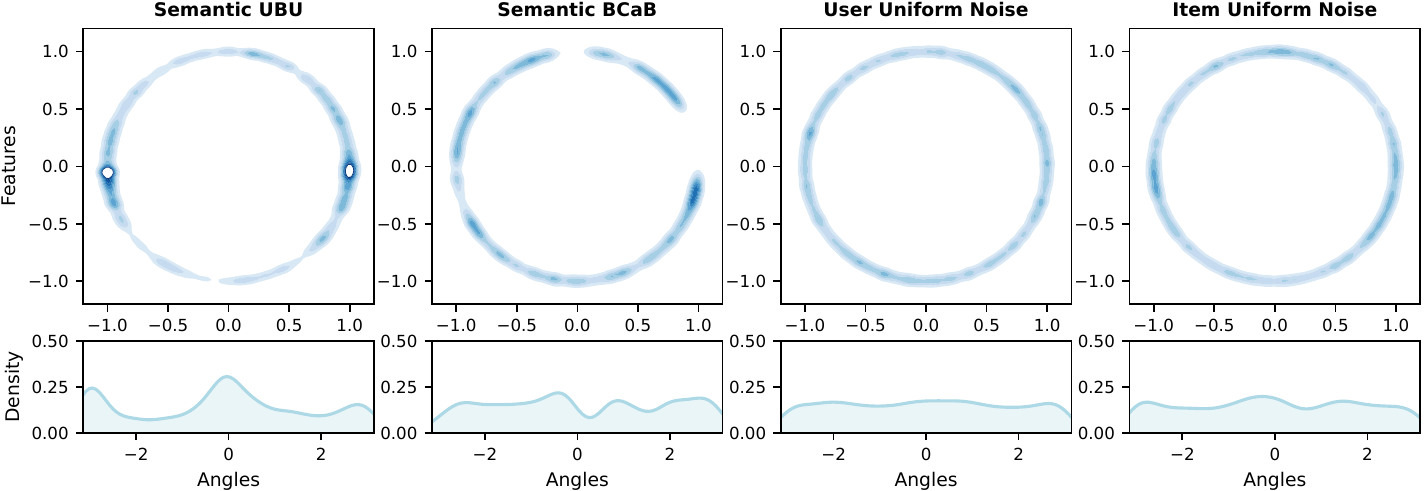}}
    \end{minipage}
    \\
    \begin{minipage}[b]{\linewidth} 
        \centering
        \subfloat[\small Distribution of representations on Douban Movie]{\includegraphics[width=\linewidth]{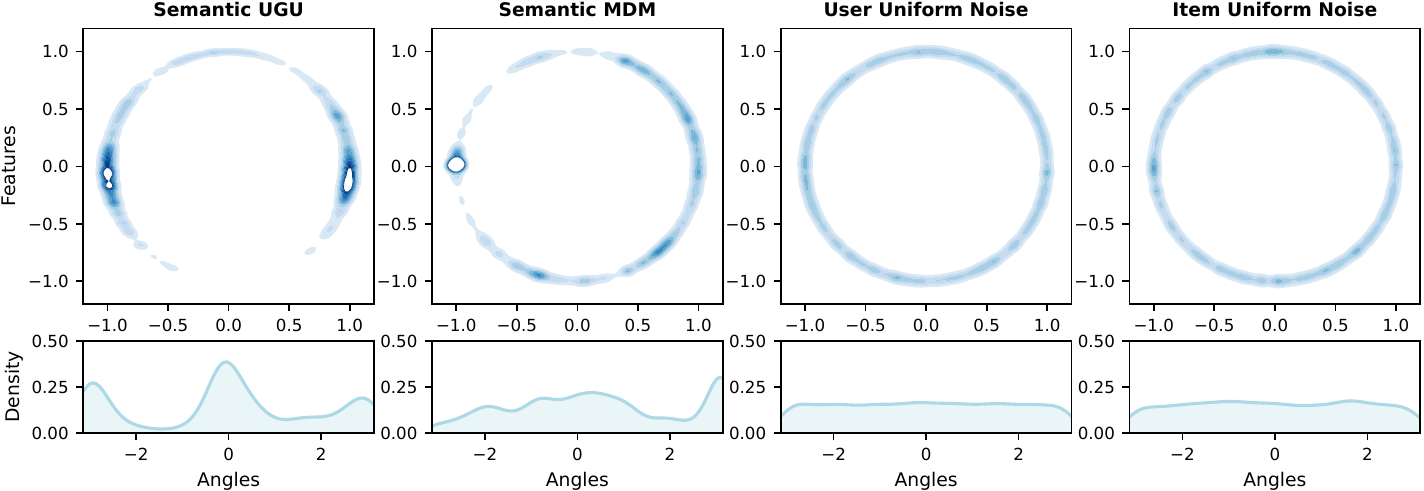}}
    \end{minipage}
    \caption{IHGCL's intents augmentation and SimGCL's uniform noise. We plot feature distributions with Gaussian kernel density estimation (KDE) \cite{2020kde} (the darker the color is, the more points fall in that area.) and KDE on angles.}
    \label{fig7:vis}
    \vspace{-0.5cm}
\end{figure}

\begin{figure}[ht]
    \centering
    \includegraphics[width=\linewidth]{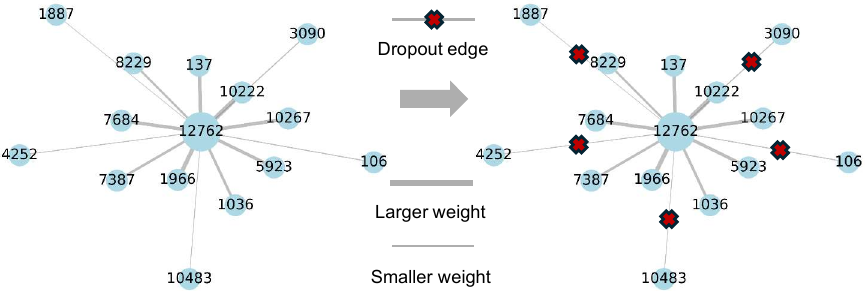}
    \caption{Meta-path UU on Douban Book. The left shows the original interaction graph of user 12762 before training, while the right is graph constrained by the information bottleneck.}
    \label{fig8:case}
    \vspace{-0.5cm}
\end{figure}

\subsection{Explainability and Visualization (RQ2)}\label{3.3}
In this section, we illustrate how intent enhances our CL-based recommendation, and experiments demonstrate that it provides more personalized recommendations for users with sparse interactions.
\subsubsection{\textbf{Comparisons w.r.t. Data Sparsity.}}
Existing models often struggle to provide reasonable recommendation for users with fewer interactions, which is one of the critical factors hindering performance. To verify IHGCL's ability to explore the intents of these users, we conducted sparsity tests on four datasets in Figure \ref{fig6:sparsity}. Following the settings \cite{2019ngcf} and \cite{zhang2023revisiting}, we divided users into four groups based on the scale of their interaction, ensuring that the total number of interactions in each group was roughly equal. Taking the sparsest dataset `Yelp' as an example, the first group contains 6,721 users with interactions that do not exceed 22. In other words, 75\% of users in the test set exhibit very sparse interactions. Figure \ref{fig6:sparsity} demonstrates that as the interaction scale increases, the performance of all methods improves significantly, highlighting that more interaction data is crucial for recommendation. IHGCL achieved performance improvements in the sparsest user group by 4.21\%, 10.19\%, 4.60\%, and 4.05\% (relative to SimGCL \cite{2022simgcl}), proving that modeling heterogeneous information to capture the intents of users and items is crucial for alleviating sparsity issue.

\begin{figure*}[t]
    \centering
    \begin{minipage}[b]{0.46\linewidth}
        \centering
        \subfloat[\small Amazon]{\includegraphics[width=\linewidth]{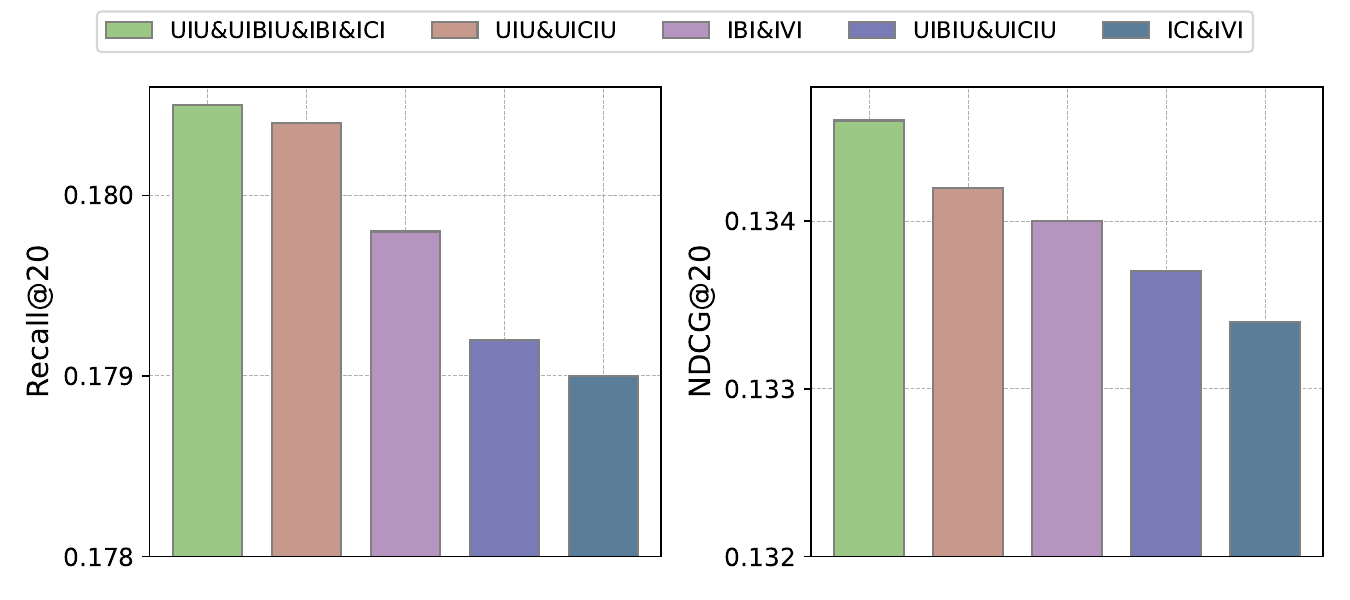}}
    \end{minipage}
    \begin{minipage}[b]{0.46\linewidth} 
        \centering
        \subfloat[\small Yelp]{\includegraphics[width=\linewidth]{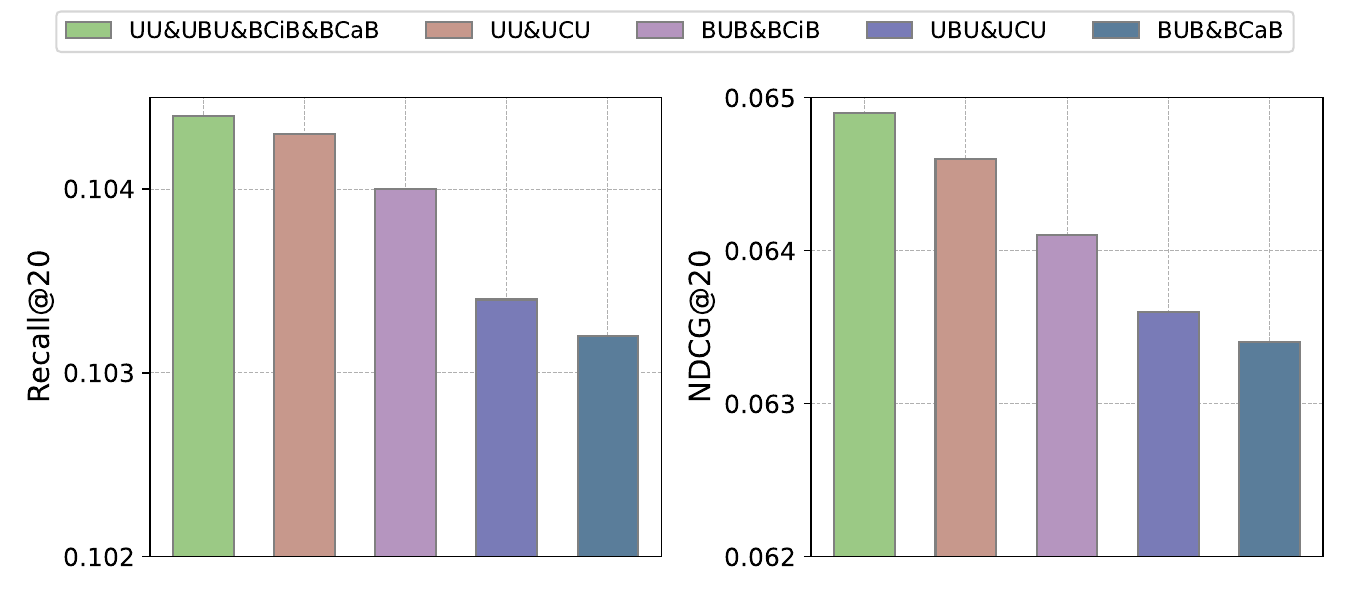}}
    \end{minipage}
    \\
    \begin{minipage}[b]{0.46\linewidth}
        \centering
        \subfloat[\small Douban Book]{\includegraphics[width=\linewidth]{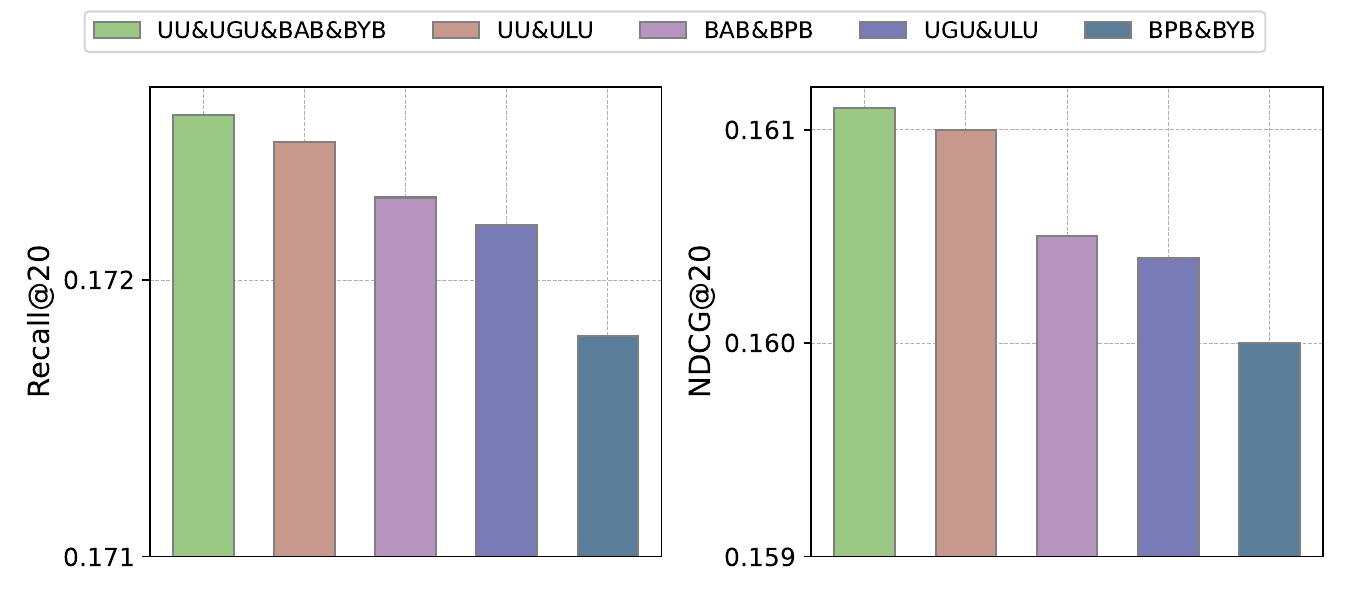}}
    \end{minipage}
    \begin{minipage}[b]{0.46\linewidth} 
        \centering
        \subfloat[\small Movielens]{\includegraphics[width=\linewidth]{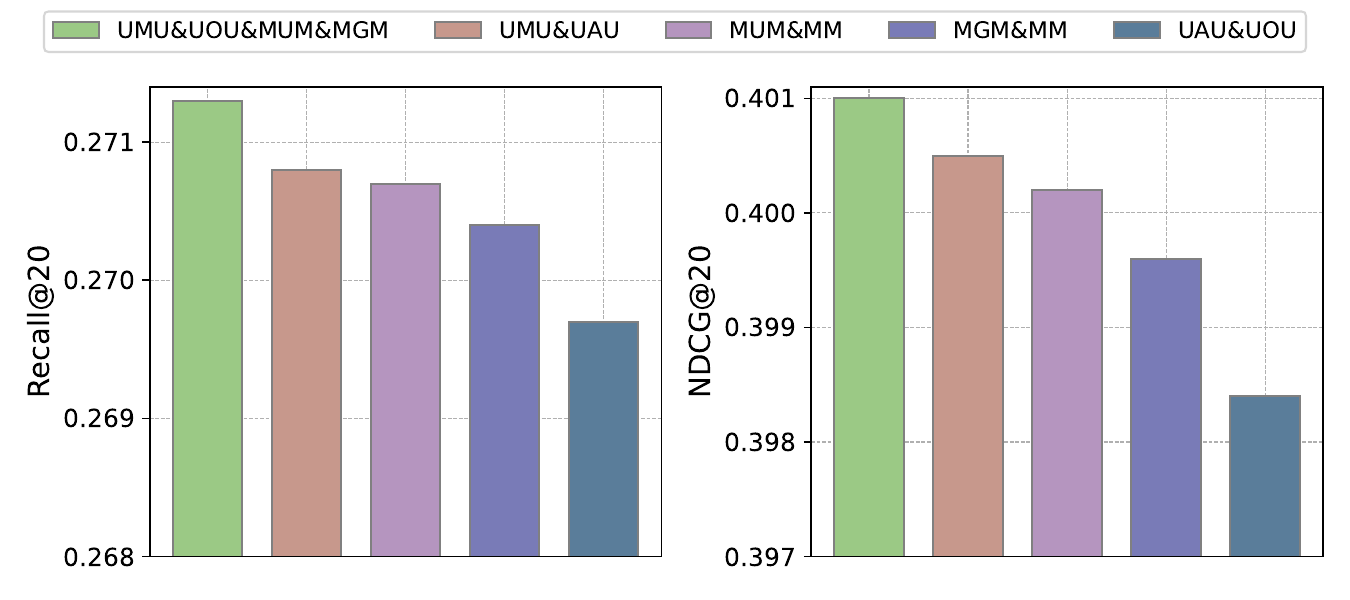}}
    \end{minipage}
    \caption{The impact of different meta-path combinations. The green bar represents the optimal meta-path we used, while the others show the replaced meta-paths. \textbf{The minimum value on the $y$-axis is significantly better than strongest baselines.}}
    \label{fig9:mp_selection}
\end{figure*}

\subsubsection{\textbf{Modeling Heterogeneous Information.}}
According to Eq. \ref{eq22}, we utilize t-SNE \cite{2008tsne} to visualize the intents of the modeled users and items. As shown in Figure \ref{fig7:vis}, following the setting \cite{2022simgcl}, we input the trained embeddings into BAE to obtain intent-based representations of users and items. For example, in the Yelp dataset, , the two columns on the left represent features obtained by encoding intent behind meta-paths, while the two columns on the right represent uniform noise added at each layer by SimGCL. We observe that, compared to the uniform noise in SimGCL, the users and items in IHGCL exhibit an apparent deviation. These deviation guide the CL-based gradient, leading to parameter updates in the intended direction. We explain this phenomenon in Figure \ref{fig5:analysis} (a) and utilize these preferences to improve the representations of users and items. Such visualization results demonstrate that modeling heterogeneous information facilitates the understanding of the intents of users and items, as evidenced by the sparsity test.

\subsubsection{\textbf{Case Study of BAE}}
As shown the left of Figure \ref{fig8:case}, before training begins, user 12762 is connected to other users through the meta-path UU. After training, we obtain the re-parameterized edge weights by one iteration of BAE using the user embeddings. The thickness of the edges in the graph represents the magnitude of these weights. On the right side, BAE reconstructs the graph based on a threshold ($cf$.Eq. \ref{eq10}), removing some connections to achieve a minimized graph based on the information bottleneck principle. Overall, the superior recommendation performance and the graph structure minimization process demonstrate that BAE can effectively mitigate noise issues, addressing \textbf{CH2}.

\begin{table}[t]
    \captionsetup{justification=centering}
    \caption{Performance w.r.t. Meta-path Number.}
    \begin{adjustbox}{width=0.48\textwidth}
    \begin{tabular}{ccccccc}
    \toprule[1pt]
    \textbf{Meta-path} & \multicolumn{2}{c}{ \textbf{Douban Book} } & \multicolumn{2}{c}{ \textbf{Yelp} } & \multicolumn{2}{c}{ \textbf{Movielens} } \\
    \cmidrule{2-7} \textbf{Number} & R@20 & N@20 & R@20 & N@20 & R@20 & N@20 \\
    \midrule \textbf{2-MP}
    & 0.1726 & 0.1611 & 0.1044 & 0.0649 & \textbf{0.2713} & \textbf{0.4010} \\
    \midrule \textbf{3-MP} 
    & \textbf{0.1734} & \textbf{0.1620} & 0.1049 & 0.0653 & 0.2700 & 0.3992 \\
    \midrule \textbf{4-MP}
    & 0.1721 & 0.1606 & \textbf{0.1050} & \textbf{0.0655} & 0.2688 & 0.3985 \\
    \bottomrule[1pt]
    \end{tabular}
    \end{adjustbox}
    \label{table3:mmp}
    \vspace{-0.5cm}
\end{table}

\subsection{In-depth Analysis of Meta-paths (RQ3)}
In Section \ref{sec2.1}, we select two user and item subgraphs based on meta-paths for the model. To study the effect of incorporating multiple intents into our model, we also perform experiments to explore the following two questions:
\subsubsection{\textbf{Selection Impact with Two Meta-paths}}
CL-based recommendation methods typically construct two augmented views for alignment to mine consistency information. We follow this paradigm \cite{2021sgl, 2022simgcl}, so in the experiments, we select two intents for users and items, respectively. In Figure \ref{fig9:mp_selection}, we examine the impact of different meta-path combinations on model performance across four datasets. For example, in the Movielens dataset, the second brown bar labeled `UMU\&UAU' indicates the replacement of the path `UOU' in the optimal combination with `UAU'. We ranked these combinations in descending order and observed that: i) The meta-path UU (i.e., social intent) compared with other intents often achieves better recommendation accuracy. We believe this is due to the positive role of social context and network homogeneity in shaping user influence. ii) From the coordinate axis, it can be seen that the selection of intents can be diverse and does not significantly impact the model's performance. The lowest values across all datasets are still better than the baseline in Table \ref{table2:results}, indicating that IHGCL can adaptively capture useful information during the alignment of intents.

\subsubsection{\textbf{Generalization Ability with More Meta-paths}}
To explore the impact of aligning more intents on recommendation accuracy, we compared multiple views pairwise to generate more supervision signals. As shown in Table \ref{table3:mmp}, we designed three variants of the model on three datasets: 2-MP is the existing model, 3-MP adds one user and one item meta-path to the original, and so on. It can be observed that with the increase in contrastive views (i.e., the addition of more meta-paths), the model performance do not significantly improve, and even declined. We believe the information in these views is limited, and over-mining may lead to an information bottleneck \cite{2022cgi}. That is, the useful information that can be provided by the views has already been fully utilized. Adding more views or adjusting the model structure is unlikely to yield more useful information, thus limiting performance improvement. Moreover, this slight increase leads to a significant rise in complexity \cite{2022crosscbr}, hence we do not adopt this strategy.
\begin{table*}[t]
    \captionsetup{justification=centering}
    \caption{Ablation studies of IHGCL on all datasets w.r.t. Recall@20 and NDCG@20.}
    \begin{adjustbox}{width=\textwidth}
    \begin{NiceTabular}{l|cc|cc|cc|cc|cc|cc}
    \toprule[1pt] Datasets & \multicolumn{2}{|c|}{ Last.fm } & \multicolumn{2}{c|}{ Amazon } & \multicolumn{2}{c}{ Yelp } & \multicolumn{2}{c}{ Douban Book } & \multicolumn{2}{c}{ Movielens } & \multicolumn{2}{c}{ Douban Movie } \\
    \midrule Metrics & R@20 & N@20 & R@20 & N@20 & R@20 & N@20 & R@20 & N@20 & R@20 & N@20 & R@20 & N@20 \\
    \midrule 
    w/o DCL & 0.2690 & 0.2657 & 0.1687 & 0.1237 & 0.0949 & 0.0580 & 0.1568 & 0.1451 & 0.2557 & 0.3882 & 0.1898 & 0.2124 \\
    w/o BAE & 0.2754 & 0.2725 & 0.1742 & 0.1286 & 0.0952 & 0.0583 & 0.1589 & 0.1480 & 0.2623 & 0.3913 & 0.1950 & 0.2138 \\
    w/o IB  & 0.2787 & 0.2782 & 0.1781 & 0.1320 & 0.1010 & 0.0629 & 0.1714 & 0.1584 & 0.2687 & 0.4001 & 0.1966 & 0.2141 \\ 
    w/o ICL & 0.2769 & 0.2766 & 0.1773 & 0.1313 & 0.1021 & 0.0632 & 0.1696 & 0.1593 & 0.2685 & 0.3998 & 0.1956 & 0.2121\\
    w/o IICL & 0.2677 & 0.2638 & 0.1713 & 0.1262 & 0.0961 & 0.0586 & 0.1543 & 0.1434 & 0.2604 & 0.3926 & 0.1921 & 0.2154\\
    \midrule IHGCL & $\mathbf{0.2824}$ & $\mathbf{0.2815}$  & $\mathbf{0.1795}$ & $\mathbf{0.1346}$ & $\mathbf{0.1044}$ & $\mathbf{0.0649}$ & $\mathbf{0.1726}$ & $\mathbf{0.1611}$ & $\mathbf{0.2713}$ & $\mathbf{0.4010}$ & $\mathbf{0.2003}$ & $\mathbf{0.2195}$ \\
    \bottomrule[1pt]
    \end{NiceTabular}
    \end{adjustbox}
    \label{table4:ablation}
\end{table*}

\subsection{Ablation Study (RQ4)}
In this section, we validate the effectiveness of the key components in our IHGCL and provide possible explanations regarding the results. Here is our explanation of variants for several models.
\begin{itemize}[leftmargin=*]
\item $\text{IHGCL}_\text{w/o DCL}$: remove the dual contrastive learning;
\item $\text{IHGCL}_\text{w/o BAE}$: remove the bottlenecked autoencoder, and use a GCN \cite{2020lightgcn} encoder directly recommendation;
\item $\text{IHGCL}_\text{w/o IB}$: remove the information bottleneck in BAE;
\item $\text{IHGCL}_\text{w/o ICL}$: remove the intent-intent contrastive Learning in DCL but keep intent-interaction contrastive Learning;
\item $\text{IHGCL}_\text{w/o IICL}$: remove the intent-interaction contrastive Learning in DCL.
\end{itemize}

Table \ref{table4:ablation} shows the experimental results of all variants on all datasets, and we have the following findings: 

The experimental results confirm that the DCL module significantly contributes to improving recommendation performance, emphasizing the necessity of leveraging heterogeneous information to model user and item intents from a contrastive learning perspective. In contrast, the BAE module holds secondary importance, as it effectively alleviates noise issues by helping the model establish reliable preferences and capture forge semantic connections.
Subsequently, we further dissect the DCL into ICL and IICL to investigate their roles. It is noteworthy that IICL serves as the principal loss function. This suggests that intent-interaction contrast through aligned heterogeneous information significantly enhances intents, while intent-intent contrast regulates intent uniformity. Finally, the information bottleneck, as a crucial component of BAE, imposes constraints on the denoising objective of the autoencoder, thereby validating the effectiveness of the denoising strategy.

\begin{figure}[t]
    \centering
    \begin{minipage}[b]{0.45\linewidth}
        \centering
        \subfloat[\small Recall w.r.t. coefficient]{\includegraphics[width=\linewidth]{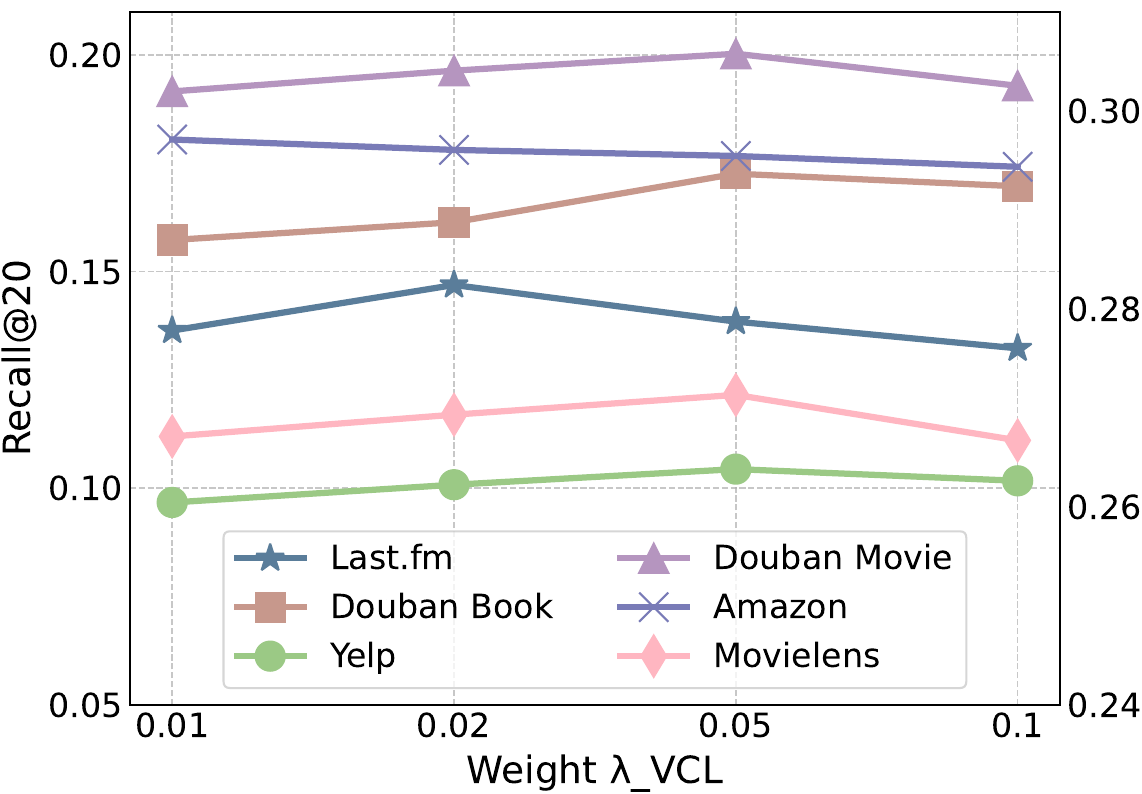}}
    \end{minipage}
    \begin{minipage}[b]{0.45\linewidth} 
        \centering
        \subfloat[\small NDCG w.r.t. coefficient]{\includegraphics[width=\linewidth]{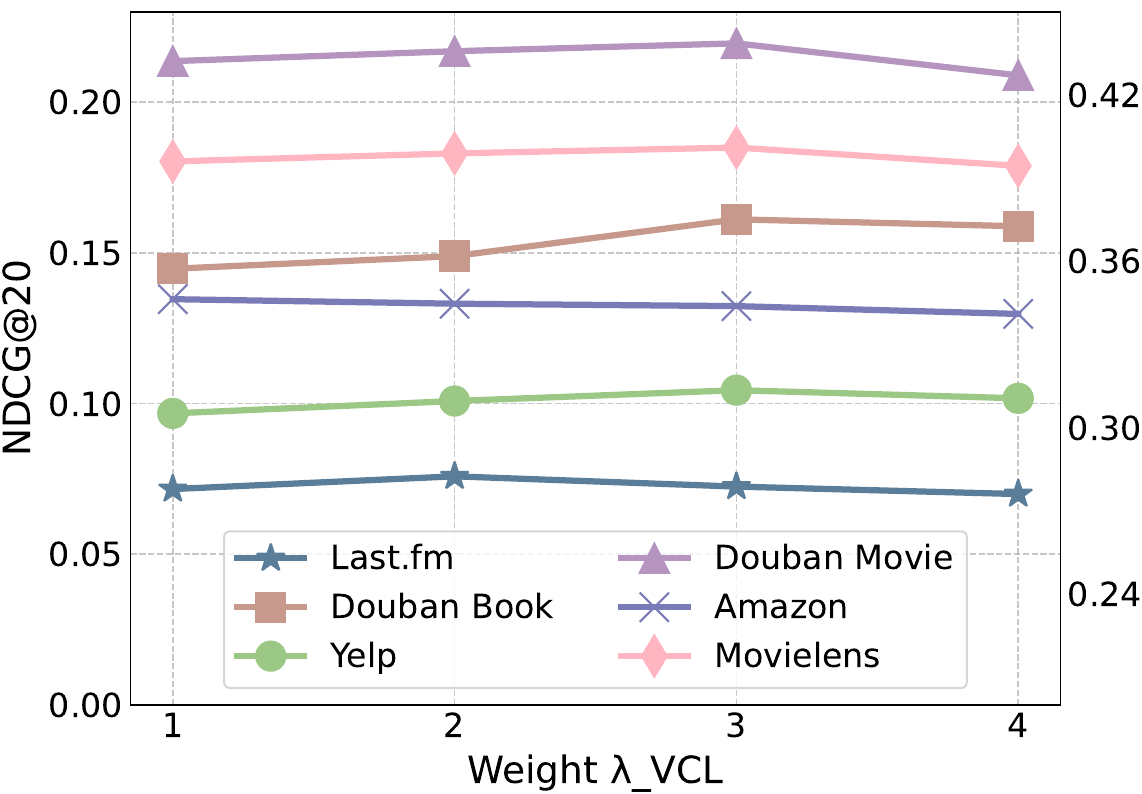}}
    \end{minipage}
    \\
    \begin{minipage}[b]{0.45\linewidth}
        \centering
        \subfloat[\small Recall w.r.t. GCN layer]{\includegraphics[width=\linewidth]{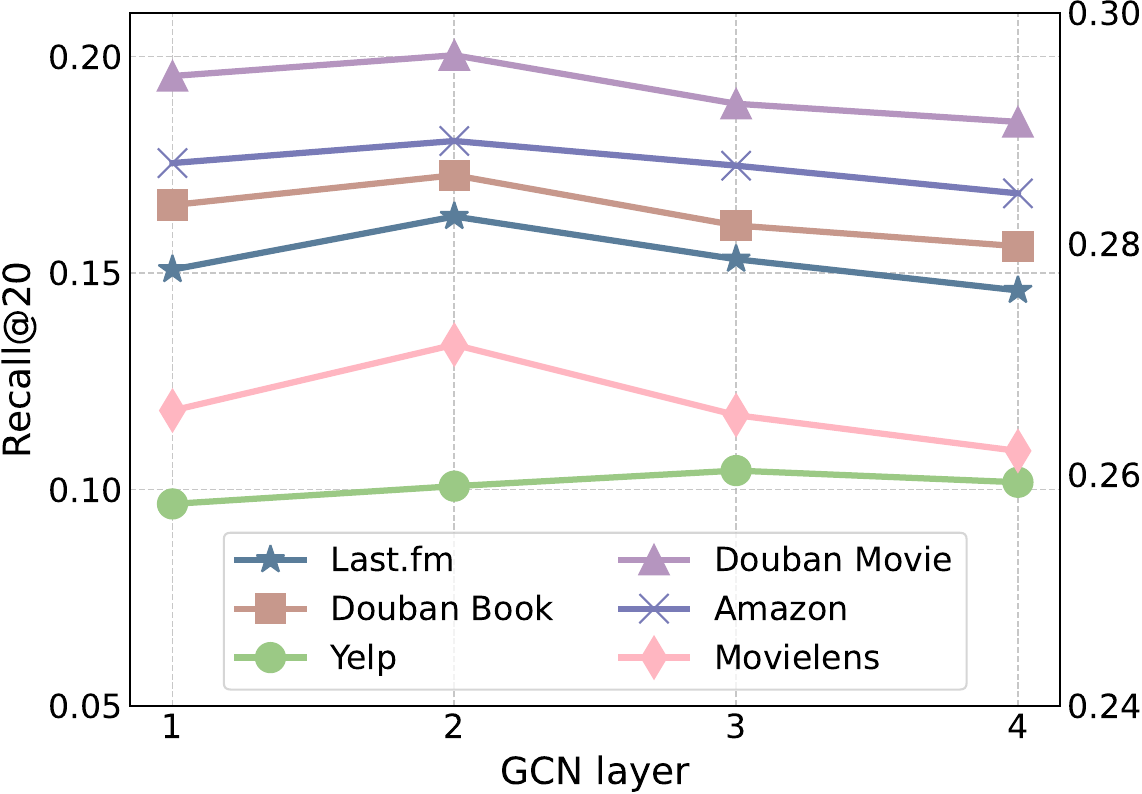}}
    \end{minipage}
    \begin{minipage}[b]{0.45\linewidth} 
        \centering
        \subfloat[\small NDCG w.r.t. GCN layer]{\includegraphics[width=\linewidth]{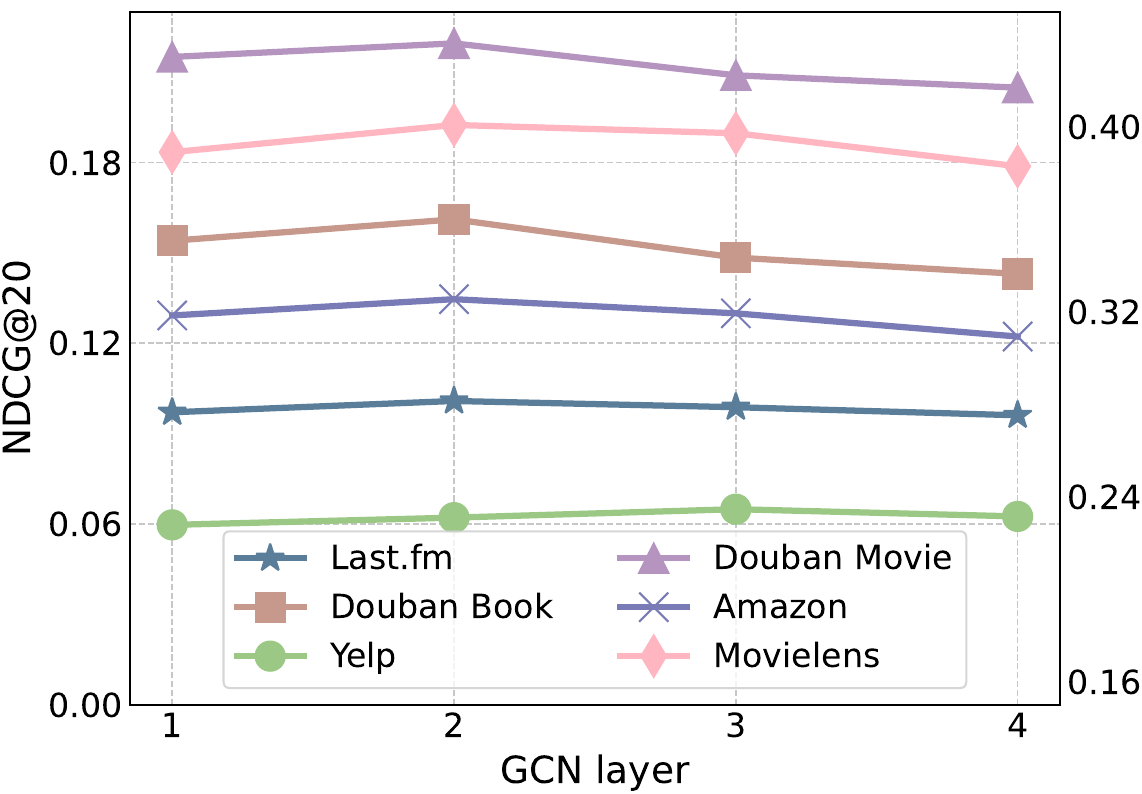}}
    \end{minipage}
    \caption{Performance comparison w.r.t. IICL coefficients and GCN layers. The axes on the right belong to the Last.fm and Movielens datasets and the right belong to other datasets.}
    \label{fig11:VCL&GCN}
\end{figure}

\begin{figure}[t]
    \centering
    \begin{minipage}[b]{0.32\linewidth}
        \centering
        \subfloat[\small Last.fm]{\includegraphics[width=\linewidth]{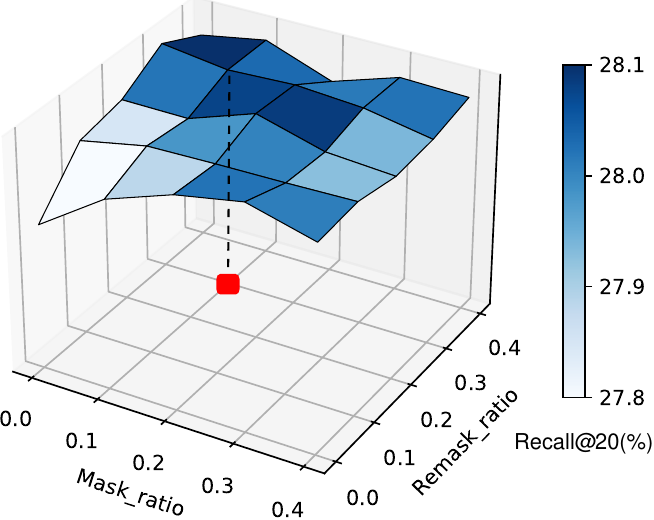}}
    \end{minipage}
    \begin{minipage}[b]{0.32\linewidth}
        \centering
        \subfloat[\small Douban Book]{\includegraphics[width=\linewidth]{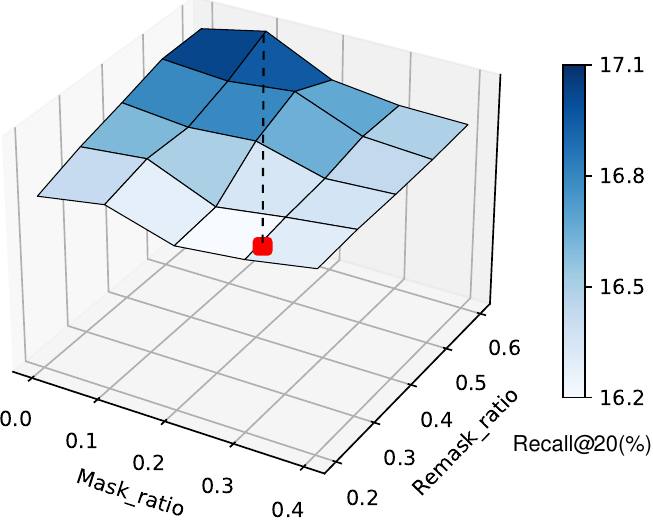}}
    \end{minipage}
    \begin{minipage}[b]{0.32\linewidth}
        \centering
        \subfloat[\small Yelp]{\includegraphics[width=\linewidth]{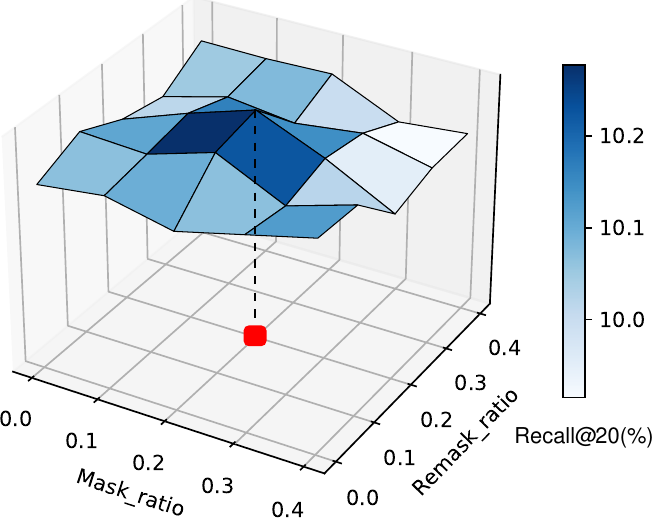}}
    \end{minipage}
    \\
    \begin{minipage}[b]{0.32\linewidth}
        \centering
        \subfloat[\small Douban Movie]{\includegraphics[width=\linewidth]{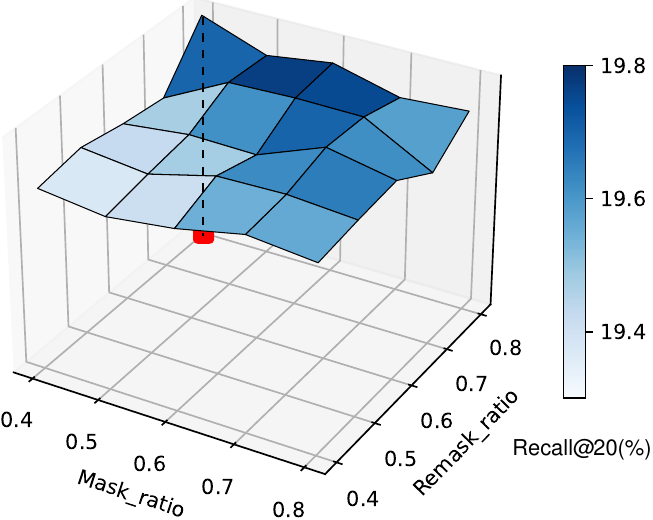}}
    \end{minipage}
    \begin{minipage}[b]{0.32\linewidth}
        \centering
        \subfloat[\small Movielens]{\includegraphics[width=\linewidth]{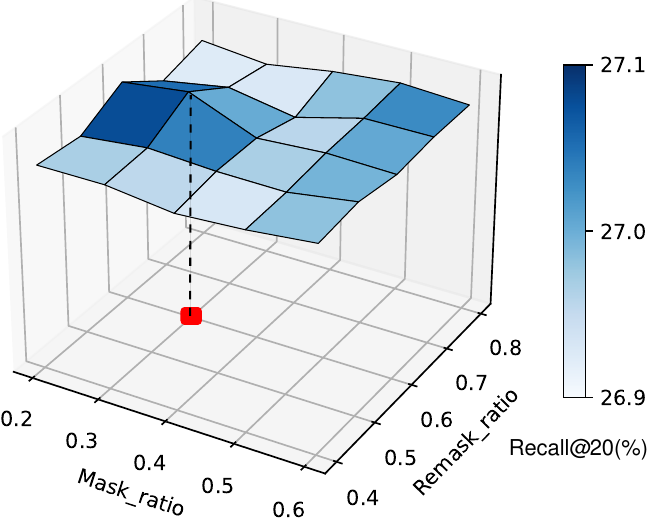}}
    \end{minipage}
    \begin{minipage}[b]{0.32\linewidth}
        \centering
        \subfloat[\small Amazon]{\includegraphics[width=\linewidth]{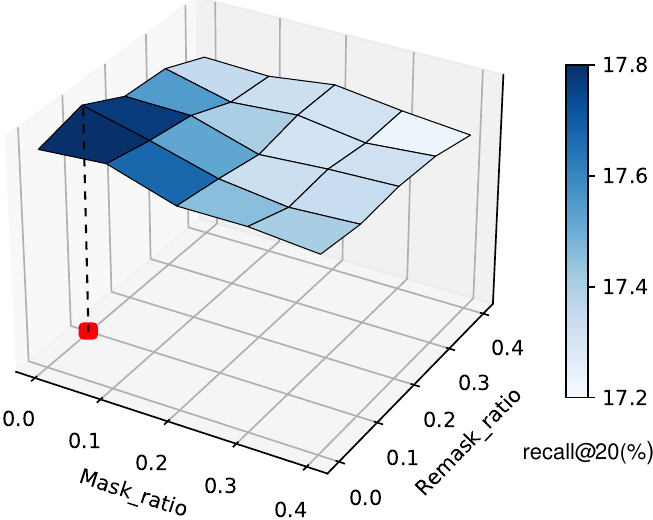}}
    \end{minipage}
    \caption{Performance comparison of different mask ratio.}
    \label{fig12:mask_ratio}
\end{figure}

\subsection{Hyperparameter Sensitivity (RQ5)}
In this study, we explored key hyperparameters' impact on our model. We investigated the effects of different losses, mask ratio in BAE and GCN layer for main task. 
\begin{itemize}[leftmargin=*]
\item \textbf{IICL coefficient.}
Our ablation studies indicated that the IICL loss exerts the most significant impact on the experimental results. To investigate the model's sensitivity to the strength of control in contrastive learning using the loss, we analyzed its impact on model performance within the range $\left({0.01, 0.02, 0.05, 0.1}\right)$ as illustrated in Figure \ref{fig11:VCL&GCN} (a) and (b). It was observed that most datasets achieved optimal performance at 0.05 (or lower), and higher values of $\lambda_{\text{IICL}}$ tended to overly emphasize the contrastive optimization loss.
\item \textbf{GCN layer.}
The results from Figures \ref{fig11:VCL&GCN} (c) and (d) suggest that when using two GCN layers in IHGCL, the model can effectively integrate the intents of users and items into the main task. However, stacking more GCN layers may cause noise issues, with this noise propagating to farther nodes through iterations, ultimately resulting in degradation.
\item \textbf{Mask ratio.}
In Section \ref{2.3.2}, we employed a masked autoencoder with dual mask, where the mask ratio plays a crucial role in eliminating noise from heterogeneous information. The red points indicated in Figure \ref{fig12:mask_ratio} represent the optimal parameter values, and most datasets exhibit a prominent peak. We can observe that larger-scale datasets typically require a higher mask ratio for denoising, thus confirming the conclusion that semantic information contains a significant amount of noise.
\end{itemize}

\section{RELATED WORK}
\subsection{Contrastive Learning for Recommendation}
Contrastive learning \cite{2020moco,2020simclr} has emerged as one of the paradigms in self-supervised learning, which aims to learn representation invariants by pulling positive samples closer and pushing negative samples apart. Recently, some studies \cite{2021sgl, 2022ncl, 2022simgcl, 2023hgcl, 2022directau, 2023lightgcl} have explored the application of contrastive learning in recommender systems. These methods leverage various embedding contrasts to generate effective self-supervised signals to alleviate the data sparsity issue. In particular, SGL \cite{2021sgl} designs three graph structure augmentation methods to construct contrastive views and maximizes the consistency between views to enhance recommendation. NCL \cite{2022ncl} explicitly models the semantic information brought by user (or item) relationships and realizes contrast between nodes of the same type through the outputs of different layers of GNN. SimGCL \cite{2022simgcl} explores the effectiveness of graph structure augmentation and proposes a simple paradigm of noise-enhanced embedding for contrast. DirectAU \cite{2022directau} analyzes the alignment and uniformity properties of contrastive learning and effectively enhances the recommendation performance by using loss functions that optimize these two objectives. LightGCL \cite{2023lightgcl} uses singular value decomposition (SVD) to construct lightweight contrastive views for recommendation.

\subsection{Heterogeneous Graph Neural Networks}
Heterogeneous graphs (HGs) have unique advantages in modeling complex relationships in the real world. In recent years, Graph Neural Networks (GNNs) \cite{2022graphmae, 2020gpt-GNN} have made significant progress in various fields, and inspired by this, Heterogeneous Graph Neural Networks (HGNNs) \cite{2019han, 2019hetgnn, 2020magnn, 2021heco, 2023hgcml, 2023hgmae} combine rich semantics indicated by different meta-paths with GNNs, achieving success in more complex graph structure domains. Specifically, these HGNNs can be classified into semi-supervised and self-supervised. Semi-supervised models include HAN \cite{2019han}, and MAGNN \cite{2020magnn}. HAN employs a dual-layer attention mechanism at both the node and semantic levels to adaptively capture the relations. MAGNN integrates meta-paths and adaptive mechanisms to learn node representations on heterogeneous graphs by learning meta-path attention weights at both the node and global levels. The self-supervised models include HeCo \cite{2021heco}, and HGMAE \cite{2023hgmae}. HeCo contrasts views across network schema and meta-path to achieve self-supervised augmentation. Meanwhile, HGMAE further introduces dynamic masking mechanisms and position-aware encoding to enhance the model's performance.
\subsection{Disentanglement-based Recommendation}
In general, disentanglement-based approaches are typically grounded in modeling user-item interactions by projecting them into different feature spaces \cite{2019MEIRec, 2018VAECF}. For example, MacridVAE \cite{2019MacridVAE} encodes multiple user intents with the variational autoencoders \cite{2019DGNN}. DGCF \cite{2020DGCF} learn disentangled user representations with graph neural networks. DisenHAN \cite{2020disenhan} employs meta relation decomposition and disentangled propagation layers to capture semantics. In CDR \cite{2021CDR}, a dynamic routing mechanism is designed to characterize correlations among user intentions for embedding denoising. KGIN \cite{2021kgin} proposes the notion of shared intents and captures user’s path-based intents by introducing item-side knowledge graph. Some novel efforts attempt to integrate contrastive learning into intent modeling, such as ICLRec \cite{2022ICLRec}, DCCF \cite{2023dccf}, and BIGCF \cite{2024bigcf}. DCCF \cite{2023dccf} enhances self-supervised signals by learning disentangled representations with global context. BIGCF explores the individuality and collectivity of intents behind interactions for collaborative filtering. However, these efforts do not consider the combination of fine-grained intents and contrastive learning in heterogeneous graphs. The proposed IHGCL aims to replace unreliable data augmentation with modeled intent embeddings.

\section{Conclusion}
In this paper, we considered leveraging fine-grained intents of users and items in heterogeneous graph recommendation to construct augmented views for contrastive learning, and we represented these intents from an explainable standpoint using various meta-paths. Based on this, we proposed a novel end-to-end recommendation model: Intent-guided Heterogeneous Graph Contrastive Learning (IHGCL). We introduced dual contrastive learning to incorporate the captured intents behind meta-paths into the main user-item view, forming augmented contrastive views and aligning intents across meta-paths. Additionally, to alleviate the noise issues inherent in intents, we further proposed an bottlenecked autoencoder that combines mask and information bottleneck. Finally, we conducted extensive experiments on six real-world datasets to validate the effectiveness of IHGCL.

\section*{Acknowledgment}
This work is supported by the National Natural Science Foundation of China (No. 62272001 and No. 62206002), Anhui Provincial Natural Science Foundation (2208085QF195), and Hefei Key Common Technology Project (GJ2022GX15).
\nocite{*}
\bibliographystyle{IEEEtran}
\bibliography{ref}

\end{document}